\documentclass[twocolumn,structabstract,longauth]{aa} 
\usepackage{graphicx}
\usepackage{txfonts}
\usepackage[]{natbib}
\bibpunct{(}{)}{;}{a}{}{,}
\usepackage{subfig}
\usepackage{epsf}
\usepackage{rotating}
\usepackage{graphics}
\usepackage{amssymb}
\newcommand{\corot}{\emph{CoRoT}}

\newcommand{\corottwo}{{CoRoT-2b}}

\newcommand{\corotseven}{{CoRoT-7}}
\newcommand{\corotsevenb}{{CoRoT-7b}}

\newcommand{\corots}{{\corotseven}}

\newcommand{\MJ}{M$_{Jup}$}
\newcommand{\RJ}{R$_{Jup}$}
\newcommand{\ME}{M$_{Earth}$}
\newcommand{\RE}{R$_{Earth}$}
\newcommand{\mstar}{{\rm M}$_{\star}$}
\newcommand{\rstar}{{\rm R}$_{\star}$}
\newcommand{\tstar}{{\rm T}$_{\star}$}
\newcommand{\mmplanet}{{\rm M}_{pl}}
\newcommand{\mrplanet}{{\rm R}_{pl}}
\newcommand{\mmstar}{{\rm M}_{\star}}
\newcommand{\mrstar}{{\rm R}_{\star}}

\newcommand{\mplanet}{{\rm M}$_{pl}$}
\newcommand{\rplanet}{{\rm R}$_{pl}$}
\newcommand{\tplanet}{T$_{pl}$}
\newcommand{\teff}{{\rm T}$_{\rm eff}$}
\newcommand{\logg}{log {\it g}}
\newcommand{\met}{[M/H]}
\newcommand{\kms}{km\,s$^{-1}$}

\newcommand{\vsini}{$v$\,sin\,$i$}   
\newcommand{\vrad}{$v_{\rm rad}$} 

\newcommand{\F}{{\rm F}}
\newcommand{\R}{{\rm R}}
\newcommand{\M}{{\rm M}}

\newcommand{\caii}{Ca {\sc  ii}}
\newcommand{\fei}{Fe {\sc   i}}
\newcommand{\lii}{Li {\sc   i}}
\newcommand{\harps}{\emph{HARPS}}
\newcommand{\uves}{\emph{UVES}}
\newcommand{\exodat}{\emph{Exo-Dat}}
\newcommand{\mass}{\emph{2-MASS}}
\newcommand{\usnoa}{\emph{USNO-A2}}

\newcommand{\tycho}{\emph{TYCHO}}

\begin{document}
%
%
   \title{Transiting exoplanets from the  \corot\, space mission
   \thanks{The \corot\ space mission, launched on 27 December 2006, has been developed and 
is operated by CNES, with the contribution of Austria, Belgium, Brazil, ESA, Germany, and Spain.  
First \corot\ data are available to the public from the \corot\ archive: http://idoc-corot.ias.u-psud.fr. 
The complementary observations were obtained with MegaPrime/MegaCam, a joint project of 
CFHT and CEA/DAPNIA, at the Canada-France-Hawaii Telescope (CFHT) which is operated by 
NRC in Canada, INSU-CNRS in France, and the University of Hawaii; ESO Telescopes at the La 
Silla and Paranal Observatories under programme ID 081.C-0413(C), DDT 282.C-5015; the 
IAC80 telescope operated by the Instituto de Astrof\'\i sica de
Tenerife at the Observatorio del Teide; the Isaac Newton Telescope (INT), operated on the island 
of La Palma by the Isaac Newton group in the Spanish Observatorio del Roque de Los 
Muchachos of the Instituto de Astrofisica de Canarias; and at the Anglo-Australian Telescope that 
have been funded by the Optical Infrared Coordination network (OPTICON), a major international 
collaboration supported by the Research Infrastructures Programme of the European 
Commissions Sixth Framework Programme; Radial-velocity observations were obtained with the SOPHIE spectrograph at the 1.93m telescope of Observatoire de Haute Provence, France.}
   }
   \subtitle{VIII. CoRoT-7b: the first  Super-Earth with measured radius}
\author{
         A. L\'eger \inst{1}
          \and     D. Rouan\inst{2} 
          \and    J.  Schneider\inst{3}
          \and P. Barge \inst{4}
          \and   M. Fridlund \inst{11}
          \and   B. Samuel \inst{1}  
	  \and M. Ollivier \inst{1}
          \and   E. Guenther \inst{5}
          \and   M. Deleuil \inst{4}   
          \and   H.J. Deeg \inst{6}
% ...
          }
   \institute{ 
   Institut d'Astrophysique Spatiale, UMR 8617 CNRS , bat 121, Universit\'e Paris-Sud, 
F-91405 Orsay, France  
      \email{alain.leger@ias.fr}
    \and 
  LESIA, UMR 8109 CNRS , Observatoire de Paris, UVSQ, Universit\'e Paris-Diderot, 
         5 place J. Janssen, 92195 Meudon, France
      \email{daniel.rouan@obspm.fr}
  \and  
  LUTH, UMR 8102 CNRS, Observatoire de Paris-Meudon, 5 place J. Janssen, 92195 Meudon, France  
  \and 
     Laboratoire d'Astrophysique de Marseille, UMR 6110 CNRS, Technop\^ole de Marseille-
Etoile, F-13388 Marseille cedex 13, France 
  \and 
     Th\"uringer Landessternwarte Tautenburg, Sternwarte 5, 07778 Tautenburg, Germany
  \and    
     Instituto de Astrof\'\i sica de Canarias, C. Via Lactea S/N, E-38200 La Laguna (Spain)  
\and 
Observatoire de Haute Provence, USR 2207 CNRS, OAMP, F-04870 St.Michel 
l'Observatoire, France
\and 
School of Physics, University of Exeter, Stocker Road, Exeter EX4 4QL, United Kingdom
\and 
 Institute of Planetary Research,  DLR, Rutherfordstr. 2, 12489 Berlin, Germany
\and 
Institute for Astronomy, University of Vienna, T\"urkenschanzstrasse 17, 1180 Vienna, Austria
\and 
Research and Scientific Support Department, European Space Agency, ESTEC, 2200 
Noordwijk, The Netherlands 
% ...
}

   \date{Received February 23, 2009; accepted  xxxxx}

 \abstract{}
   { We report the discovery of very shallow ($\Delta F/F \approx 3.4 10^{-4}$), periodic dips in the 
light curve of an active V = 11.7 G9V star observed by the \corot\, satellite, which we interpret as 
caused by  a transiting companion. We describe the 3-colour CoRoT data and 
complementary ground-based observations that support the planetary nature of the companion.}
   {  We used \corot\, colours information, good angular resolution ground-based photometric 
observations in- and out- of transit,  adaptive optics imaging,  near-infrared spectroscopy, and 
preliminary results from radial velocity  measurements, to test the  diluted eclipsing binary 
scenarios. The parameters of the host star were derived from optical spectra, which were then 
combined with the \corot\, light curve to derive parameters of the companion. }
   {We examined all conceivable cases of false positives carefully, and all the tests 
support the planetary hypothesis.  Blends with separation $>$ 0.40\arcsec or triple 
systems are almost excluded with a 8 $10^{-4}$ risk left.
  We conclude that, inasmuch we have been exhaustive,  we have discovered a planetary 
companion, named \corotsevenb\, , for which we derive a period of 0.853 59 $\pm$ 3 $10^{-5}$ 
day and a radius of R$_{p}$ = 1.68 $\pm$  0.09 \RE. Analysis of preliminary radial velocity  data  
yields an upper limit of 21 \ME\,  for the companion mass, supporting the finding.}
   { \corotsevenb\, is very likely the first Super-Earth with a measured radius. This object 
illustrates what will probably become a common situation with missions such as  Kepler, namely 
the need to establish the planetary origin of transits in the absence of a firm radial velocity 
detection and mass measurement.   The composition of \corotsevenb\, remains loosely 
constrained without a precise mass. A very high surface temperature on its irradiated face, $
\approx$ 1800 -- 2600 K at the substellar point, and a very low one, $\approx$ 50 K, on its dark face assuming no atmosphere, have been derived.}
 \authorrunning{A. L\'eger  et al.}
  \titlerunning{\corotseven\,  b: the first  Super-Earth with measured  radius}
   \maketitle
%
\section{Introduction}
 
  The space mission \corot\ is performing wide-field stellar photometry at ultra-high precision 
\citep{1998EM&P...81...79R,2006cosp...36.3749B}. During an
   observing run, up to 12,000 stars  can be monitored simultaneously and continuously over  150 
days of  observation.  \corot\, is thus particularly well-suited to detecting
planets with orbital periods shorter than 50 days.  Because the transit signal is proportional to the 
planet's  projected surface, the first published  \corot\,  results \citep{2008A&A...482L..17B,
2008A&A...482L..21A,2008A&A...491..889D,2008A&A...488L..43A,2008A&A...488L..47M}  
 were focused on the population of  rather massive planets, one of which has even been quoted as ``{\it 
the first  inhabitant of the brown-dwarf desert}'', with a well-defined mass (20 $\pm$ 1  \MJ) and a 
well-defined radius (1.0 $\pm$ 0.1 \RJ) \citep{2008A&A...491..889D}. 
However,  \corot\,  has the capability of detecting significantly smaller planets, and 
 analysis of the  noise on the light curves (LC hereafter) indeed shows that in many cases it is 
not far from the 
 photon noise limit (see Aigrain et al., this volume). In the same line, blind tests performed by 
different teams of the  \corot\,  consortium on  actual LCs  where transits were added  did 
confirm that the performances of  \corot\,  are such that $\approx$ 2 R$_{Earth}$ hot Super-Earth 
planets\footnote{We define a Super-Earth as a planet larger than the Earth but  without a significant hydrogen envelope, e.g. $< $ 10$^{-3}$ times the Earth mass. It 
can be either rocky or water-rich  \citep{2004Icar..169..499L,2009arXiv0902.1640G}.}  
are within reach.  Such planets on close-in orbits  should be
accessible for stars brighter than m$_{\rm V} \approx$ 13 \citep{2006ESASP1306..283A}. 
 
 The study of small hot planets is  becoming a 
major question \citep{2008PhST..130a4010M,2009A&A...493..639M,
2009A&A...496..527B} that 
has a direct  link with planetary system formation and evolution.
 In this paper, we report the discovery of the smallest transiting object detected so far around a 
main-sequence star, an object that deserves the name of Super-Earth.   

 In transit surveys, ground-based follow-up is mandatory for 
 confirming a transiting planet candidate. In the case of \corotsevenb\,\  an intensive follow-up 
campaign has been set up, including programmes of photometry, imaging, spectroscopy, and  
 radial velocity (RV), using different ground-based facilities over the world. The results of this 
campaign allow us to exclude almost all the possible false positive cases that could mimic  a 
transiting planet. Preliminary results of RV measurements are consistent with the presence of a 
low mass planet and exclude any giant planet or  stellar companion.
 
We present  the photometric analysis of the  \corot\,  data where we discovered  this shallow transit 
candidate (Sects. \ref{sec:corotLC} and \ref{sec:corot_colors}),  as  well as the photometric and 
imaging follow-up, including adaptive optics (Sect. \ref{sec:ground_photom}), 
infrared spectroscopy (Sect.  \ref{sec:spectro}), and preliminary results of the RV measurements 
(Sect.  \ref{sec:RV}) , all done in  order to secure the planetary nature of the transiting body. 
The stellar parameters  are  presented in
Sect.  \ref{sec:stellar_params} and  planetary ones in Sect.  
\ref{sec:planet_param}. Such a small and hot planet  raises several questions about its 
composition, structure, and surface temperature, as discussed in Sect.  \ref{sec:discussion}.

\section{Photometric observations with \corot}
\label{sec:corotLC}

   \begin{figure*}
   \centering
\includegraphics[width=16cm]{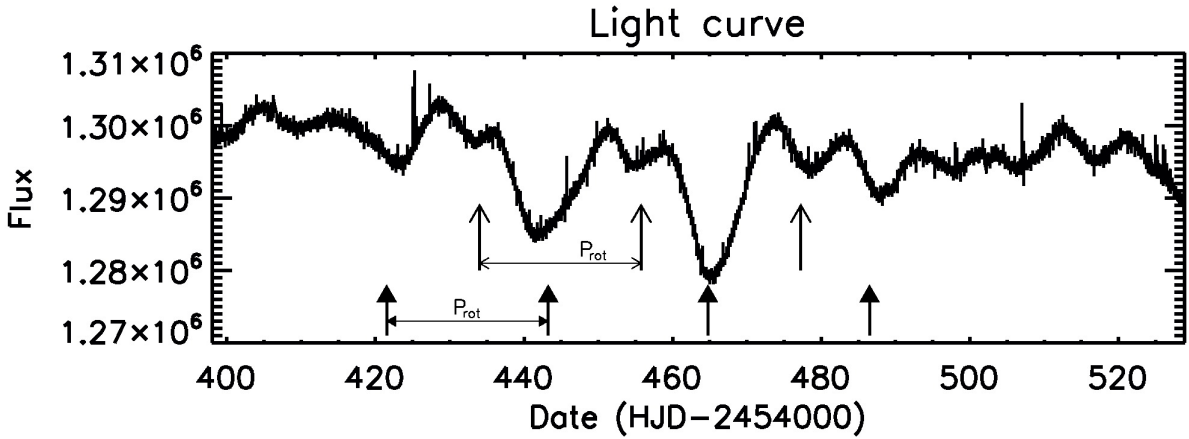}
   \caption{ LC of the target \corotseven\,  without low-frequency filtering. The stellar rotation period, 
P$_{rot}$  of 23 days can be inferred from spot induced dips, as pointed out  by arrows.
              \label{Corot_LC}}%
    \end{figure*}

   \begin{figure*}
   \centering
\includegraphics[width=15.5cm]{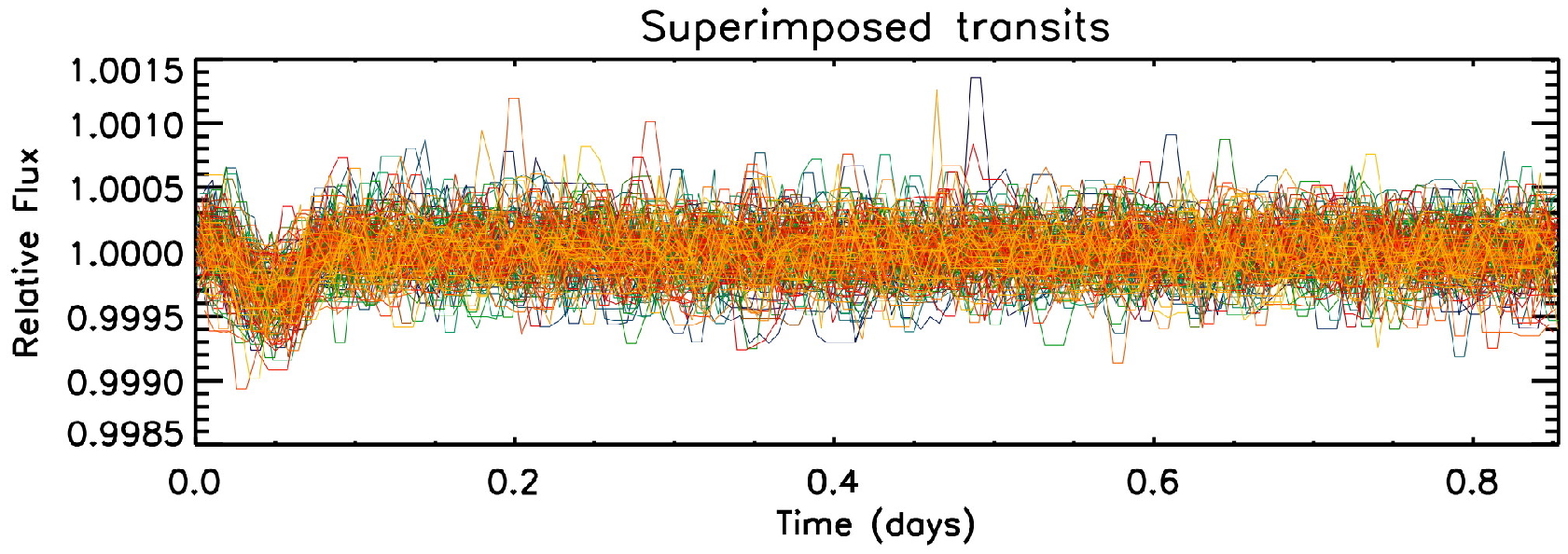}   
\includegraphics[width=16cm]{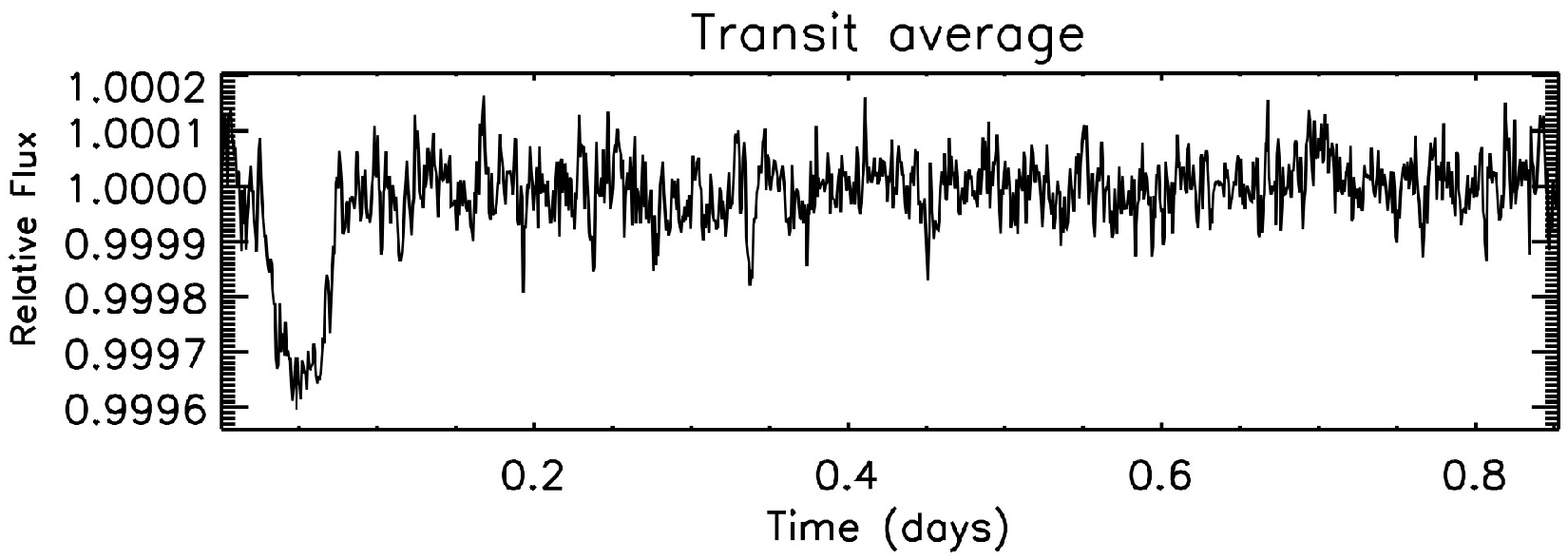}
   \caption{  Upper panel: superimposition of 153 individual segments  of the LC  divided 
according to the transit period determined by a detection algorithm after high-pass (3 times the 
transit period = 2.56 days) and low-pass (3 times the time resolution  = 3 $\times$ 512 s) filtering. Individual transits  are clearly seen when superimposed. Lower panel: mean value of the upper curves but with a 
shorter time resolution (64s) and a different low pass filtering (3 times the time resolution =  3 $
\times$ 64 s) in order to better preserve the transit shape.
              \label{LC_folded} 
              }%
    \end{figure*}
   \begin{figure*}
   \centering
\includegraphics[width=16cm]{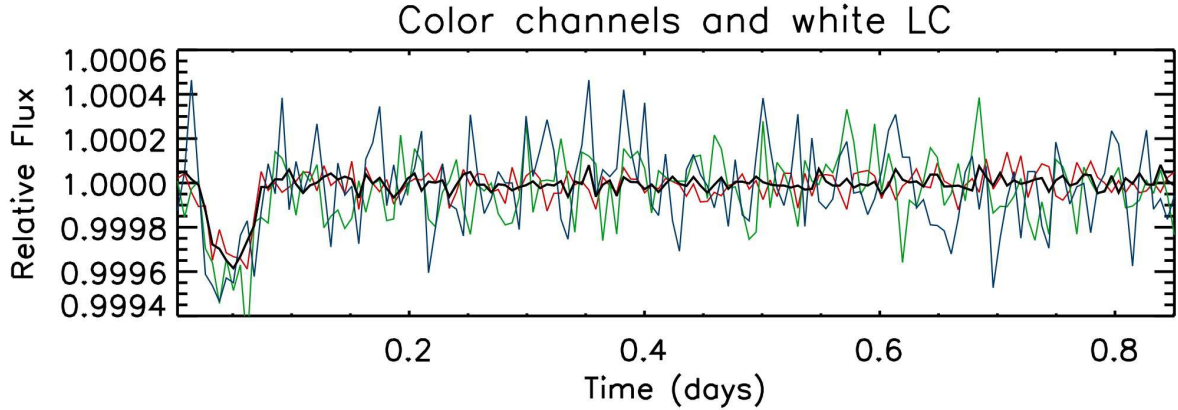}
   \caption{  
   Averaged folded LCs in the three colours provided by the \corot\, instrument, after normalization. 
Red, green and blue signals are represented with the corresponding colours, and the white 
signal, summation of the three bands prior to normalization, is in black.
              \label{LC_color}
              }%
    \end{figure*}

   \begin{figure*}
   \centering
\includegraphics[width=16cm]{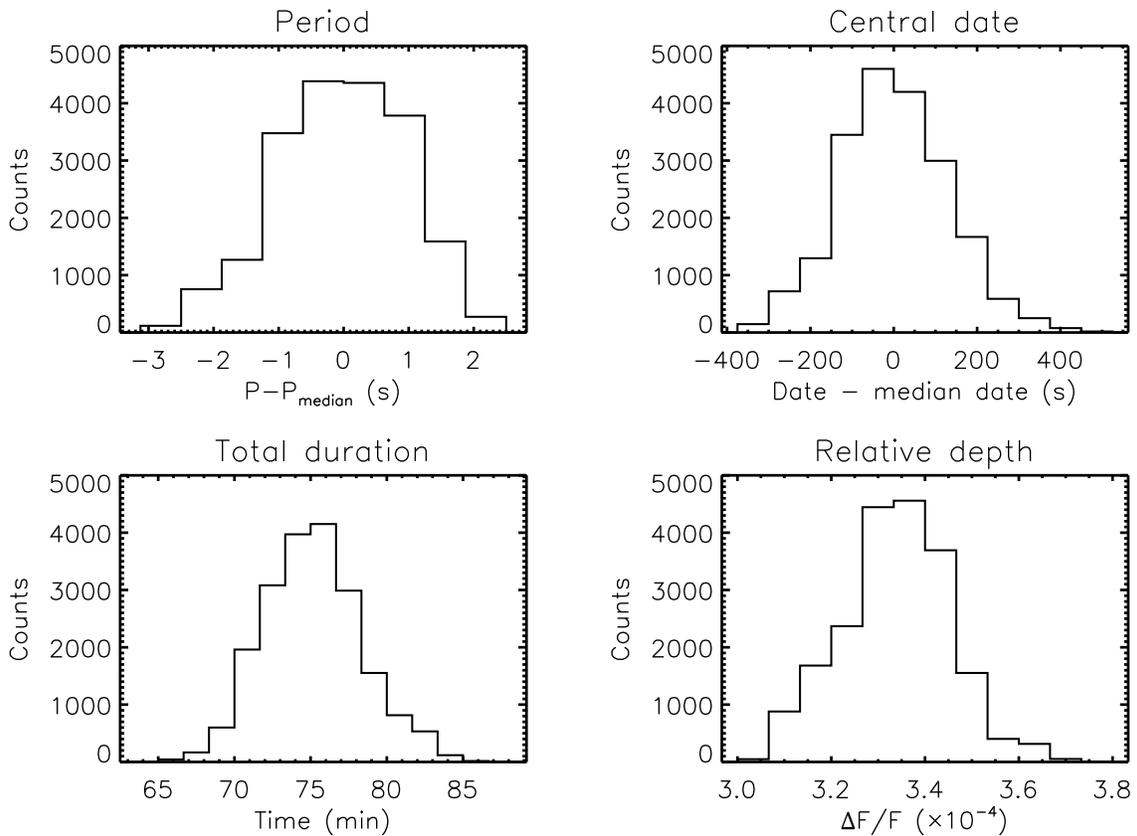}   
   \caption{  Transit parameter distributions obtained from a bootstrap method for a trapezoidal 
transit signal.
              \label{histo_bootstrap} 
              }%
    \end{figure*}

\begin{table}[h]
\caption{\label{StarID} 
\corotseven\,   IDs, coordinates, and magnitudes.}
\begin{center}{
\begin{tabular}{rcc }
\hline
\hline
 \corot\ ID & \multicolumn{2}{c}{102708694, LRa01 E2 0165} \\
\usnoa\         & \multicolumn{2}{c}{0825-03049717}\\ 
\mass\          &  \multicolumn{2}{c}{06434947-0103468} \\
\tycho\           &  \multicolumn{2}{c}{4799-1733-1}\\
RA (2000)       & \multicolumn{2}{c}{ 06:43:49.0  } \\
DEC (2000)     & \multicolumn{2}{c}{ $-$01:03:46.0 } \\
\hline
B-mag~$^{\mathrm{(a)}}$             &  12.524 & $\pm$  0.018\\
V-mag~$^{\mathrm{(a)}}$             &  11.668 &$\pm$  0.008\\
$r^\prime$-mag~$^{\mathrm{(a)}}$  &  11.378  & $\pm$ 0.008 \\
$i^\prime$-mag~$^{\mathrm{(a)}}$  &  10.924  &  $\pm$ 0.017 \\
J ~$^{\mathrm{(c)}}$     & 10.301 & $\pm$ 0.021  \\
H~$^{\mathrm{(c)}}$      & 9.880 &  $\pm$ 0.022  \\
K$_s$~$^{\mathrm{(c)}}$  & 9.806 & $\pm$ 0.019 \\
\hline
$\mu_\alpha$~$^{\mathrm{(b)}}$ & 12.9 $mas /yr$ & 1.4   \\
$\mu_\delta$~$^{\mathrm{(b)}}$ &  -4.0 $mas /yr$ & 1.5  \\
\hline
\end{tabular}}
\begin{footnotesize}
\begin{list}{}{}
\item (a) Provided by \exodat, based on observations taken at the {\sl INT} telescope. 
\item (b) From \tycho\ catalogue.
\item (c) From \mass\ catalogue.
\end{list}
\end{footnotesize}
\end{center}
\end{table}


The star \corotseven\, was observed during the first long run of  \corot\, towards the Monoceros constellation 
(anti-centre run {\sl LRa01},  the letter {\it a} indicating that the field is close to the Galactic 
anti-centre). Its ID is given in Table~\ref{StarID}, based on the {\sl Exo-Dat} database 
\citep{2009AJ....138..649D}.
Because it is one of the brightest stars monitored in this field, it was a member of the 
oversampled (32~sec) target list from the beginning of the {\sl LRa01} run. After the first 40 
days of data 
acquisition, the {\sl Alarm Mode} pipeline \citep{2006ASPC..351..307Q,2008ASPC..394..373S} 
detected the first series of transits in the star LC.

As illustrated by the whole \corot\ LC  (Fig~\ref{Corot_LC}), \corotseven\, is an active star. Its LC  
shows $\approx$ 2 \% modulations, interpreted as the effect of stellar spots driven by the stellar 
rotation and crossing the disk.  A period of  $\approx$ 23 days is inferred.

Several teams of the \corot\, exoplanet consortium have searched for transits.   Spurious 
spikes and stellar variations  at frequencies outside the range expected for planetary transits  were 
removed with low- and high-pass filters. 
Then, different detection algorithms were used 
 and 153 individual transits were  eventually detected.  In agreement with the CoRoT consortium rules, the \corotseven\, LC, with the same data pre-processing as in the present paper (v1 of data pipeline), will be accessible from 30 July 2009 
at http://idoc-corot.ias.u-psud.fr/index.jsp (select: CoRoT Public N2 data / Run LRa01 / object with 
Corot ID 102708694), so that the reader can make his or her own reduction and analysis of the 
data.

 
 In this section, we seek to estimate the main transit parameters, i.e., period, central date, ingress/
egress duration, total duration, and relative depth, using a simple trapezoidal model. We proceed 
as follows: (1) outliers (mainly due to the satellite crossing of the South Atlantic anomaly) are 
filtered out from the LC using a 7-sample running median; (2) long-term stellar activity is removed 
by subtracting a 0.854-day running median; (3) individual transits are corrected from a local linear 
trend computed on 3.75-hr windows centred on the transits but excluding the transits themselves; 
(4) a least-square fit of a trapezoidal transit signal is performed using only data inside 3.75-
hr windows centred on each transit.

Errors on the transit parameters were estimated using a procedure analogous to the bootstrap 
method described by \citet{press}, although slightly modified in order to preserve the correlation 
properties of the noise: (1) we compute a transit-free LC by subtracting our best-fit trapezoidal 
model to the data; (2) we re-insert the same transit signal at a randomly chosen phase; (3) we fit a 
trapezoidal model to the data and record the best-fit parameters; (4) steps 1-3 are repeated 
20,000 times to build histograms used as estimators of the probability distributions for every transit 
parameter (Fig. \ref{histo_bootstrap}).  Finally, the error on a given parameter is computed as the 
median absolute deviation of its distribution.
 
   \begin{figure}[h]
   \centering
\includegraphics[width=8cm]{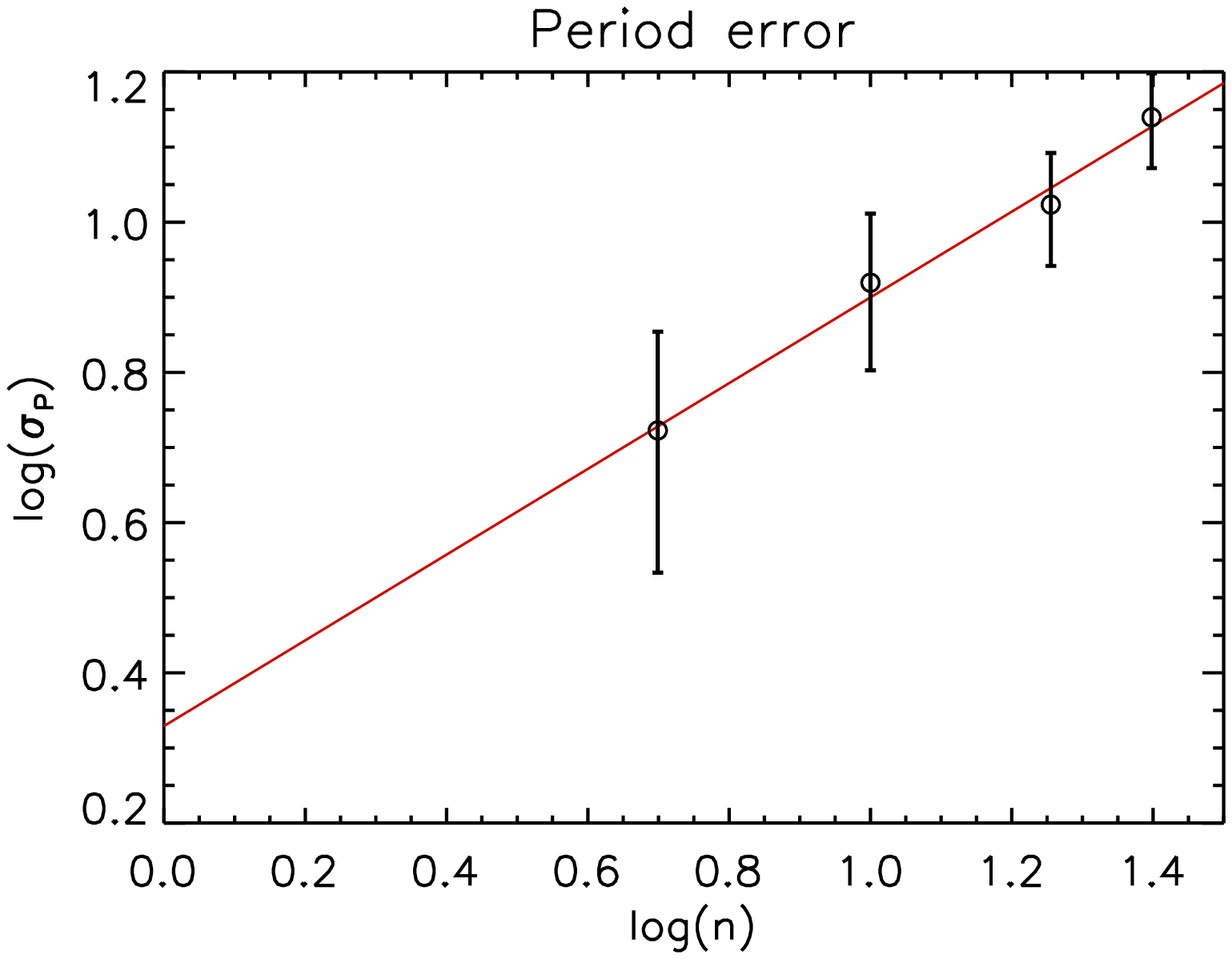}
\includegraphics[width=8cm]{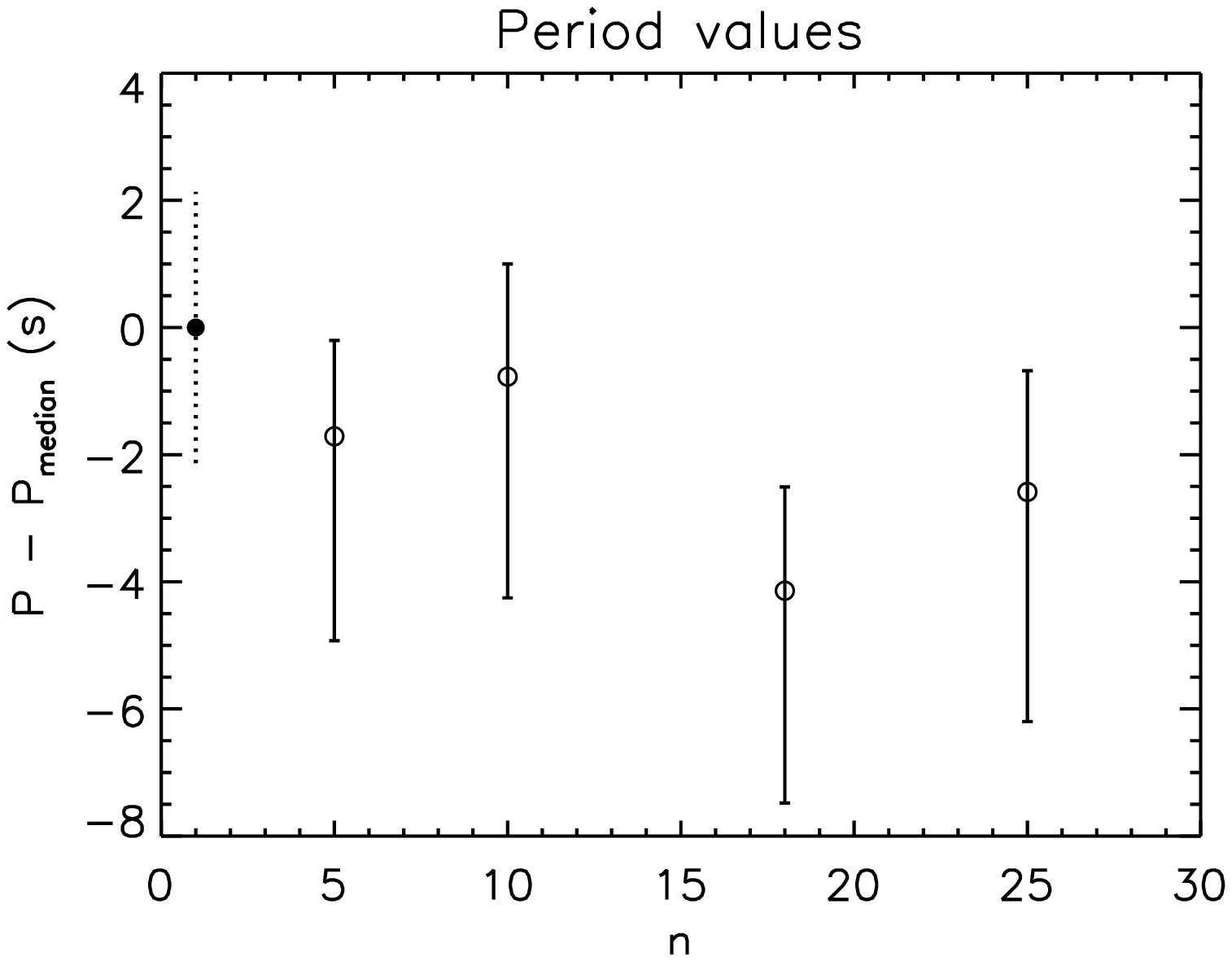}
   \caption{ 
 Period error calculation for the full LC based on a extrapolation of period error 
estimates from chopped LCs containing one transit out of n and spanning the LC total duration. 
Left: period errors as a function of n. Right: deviation from the period value $P_{median}$ yielded 
by the bootstrap.
              \label{period-error}
              }%
    \end{figure}

Because the period error yielded by the bootstrap seemed fairly small, we decided to check this 
result by carrying out a different calculation: (1) we produce 4 sets of n = 5, 10, 18, and 25 
sub-LCs, each LC having the total duration of the initial LC,  by keeping just one complete transit 
period out of 5,10,18, and 25 consecutive transits respectively; within one set, the first transit of a 
sub-LC is shifted of one with respect to the previous sub-LC; (2) for each of the 4 sets, we measure 
the period for every sub-LC with a trapezoidal least-square fit, compute the median period, the 
period standard deviation $\sigma_{P}$, and the error on the standard deviation 
(Fig. \ref{period-error}); (3) we estimate the period standard deviation for the full LC by 
extrapolating $\sigma_{P}$ for n = 1. For this purpose, we performed a least-square fit of a power
law of the form $\sigma_{P}$(n) = $\sigma_{P}$(1) $\times$ n$^{\alpha}$ and got  $\alpha$ = 0.57 
(close to the 0.5 exponent expected for uncorrelated measurements) and $\sigma_{P}$(1) = 2.1 s. 
We note that this error is a factor of 2 larger than the one obtained with the bootstrap method, so 
we conservatively chose to keep this higher value as our final estimate of the period error  (Table 
\ref{Tab_transit_param}).
 
\begin{table}[h]
\caption{\label{Tab_transit_param} 
Transit parameters and associated uncertainties, as modelled with a trapezoid.}
\begin{center}{
\begin{tabular}{ll }
\hline
period	& 0.853 585 $\pm$  0.000 024 day \\
central date (1$^{st}$ transit)	& 2 454 398.0767 $\pm$ 0.0015 HJD  \\
ingress/egress duration & 15.8 $\pm$  2.9 min \\
total duration (trapezoid) 	& 75.1 $\pm$ 3.2 min  \\
depth (trapezoid) &	3.35 10$^{-4}$ $\pm$ 0.12 10$^{-4}$ \\
\hline 
\end{tabular}}
\end{center}
\end{table}

 We finally find a period of 0.853585 $\pm$ 24 10$^{-6}$ day.  Figure \ref{LC_folded}, where all 
transits are superimposed, shows that even individual transits can be tracked down despite the 
low  S/N. The fit by a trapezoid on the average curve yields the parameters: $\tau_{23}$ = 
0.808 h, $\tau_{14}$= 1.253 h for the short and long bases of the trapezium\footnote{We use 
$\tau_{ij}$ for the parameters related to the trapezoidal fit
and T$_{ij}$ for the parameters related to the more realistic transit modelling (see Sect. 
\ref{sec:planet_param}) }, and  $\Delta F/F$ = 3.35 10$^{-4}$ $\pm$ 0.12 10$^{-4}$. This is the {\it 
faintest relative flux change that has 
been detected in transit search  photometry, up to now}.

The \corot\, camera is equipped with a low-dispersion device (a bi-prism) before the exoplanet 
CCDs \citep{2000ESASP.451..221R,1998EM&P...81...79R,2006ESASP1306..283A} that provides 
LCs in three colours, called red, green and blue, even if the band 
pass does not correspond to classical photometric filters. The corresponding phase-folded and 
averaged LCs are shown in Fig.\ref{LC_color}. The transit is observed in the three colours with 
similar relative depths, a behaviour expected for the transit of a planet in front of its star that will be used to 
assess the planet hypothesis (Sect. \ref{sec:corot_colors}).
 
\section{Photometric and imaging follow-up
}
\label{sec:ground_photom}

   \begin{figure}
   \centering
\includegraphics[width=8.5cm]{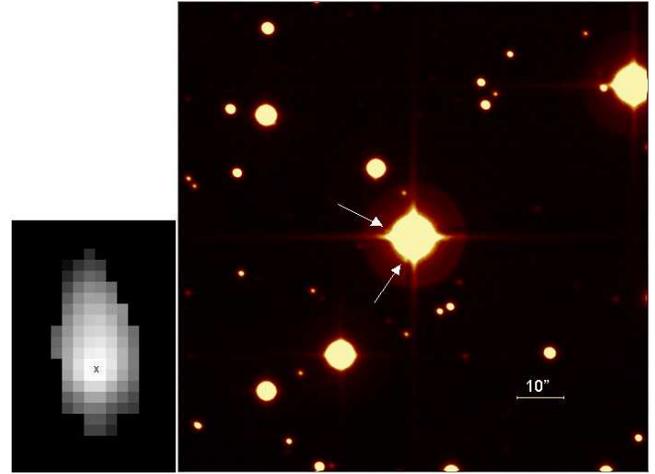}
   \caption{Shift-and-add image  of the stack of 100 exposures taken at CFHT with MEGACAM. The 
arrows  indicate the two faint stars located at $\approx$ 5 arcsec  from \corots\, (at the centre of the 
image) and about $\Delta m$ =  10 mag fainter. The size of the field shown is 1 arcmin (see scale); 
north is to the top and east to the left. The insert to the left shows the shape, size, and orientation of 
the photometric mask applied on the star \corots\, onboard  \corot\, and the X marks the position of 
the star on the mask. The grey levels corresponds to the measurement by  \corot\,  on an imagette. 
              \label{Fig_CFHTadd}}%
    \end{figure}
   \begin{figure}
   \centering
\includegraphics[width=8.5cm]{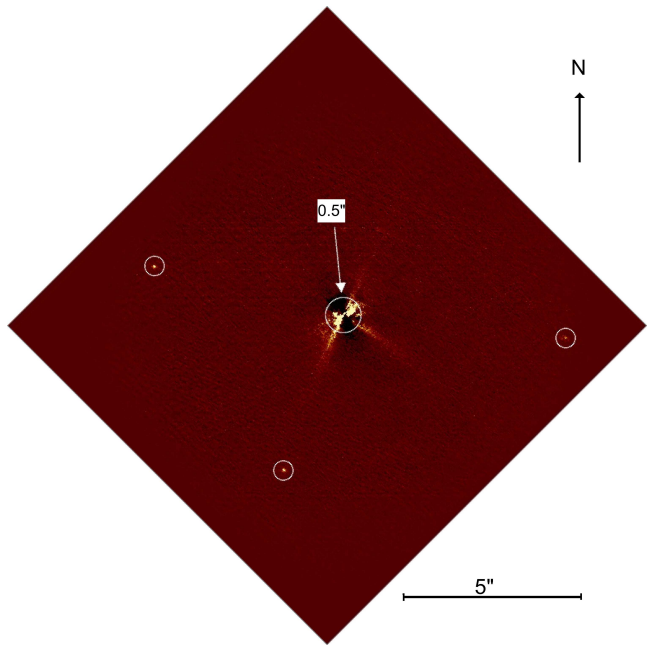}
      \caption{Final NACO image in J-band after substraction of a median \emph{PSF} and 
      de-rotation of the field.  The circles locate the three faint stars that are detected in the field, the two at 
east identified with those marked by arrows in Fig. \ref{Fig_CFHTadd}. The scale is given 
by a line of 5 arcsec length, and the central circle of 0.5 arcsec radius gives  an idea of 
the angular distance at which the presence of a faint star could not be detected close to   
\corotseven\,.   North is at top, and east to the left, as in Fig. \ref{Fig_CFHTadd}.      \label{Fig2}
}
   \end{figure}

Whenever  transits are detected in a \corot\ LC and when the candidate survives the set of tests 
performed to rule out obvious stellar systems (see Carpano et al, this volume) a ground based 
follow-up programme is initiated.  The goal is to check further for possible contaminating  eclipsing 
binaries (EBs) whose point spread function (\emph{PSF}) could fall within the \corot\ photometric 
mask. In the specific case of \corotsevenb\,\ with its shallow transits, we performed a rigorous 
complementary observational campaign  to check all conceivable  blend scenarios.
These included {\it i)} a search for photometric variations on nearby stars 
during the assumed transit;  {\it ii)}  deep  imaging, with good-to-high angular resolution,  
searching for the presence of fainter and closer contaminating stars;  {\it iii)} spectroscopic 
observations of the target at high resolution and high S/N;  {\it iv)} infrared spectroscopy, 
searching for faint low-mass companions; {\it v)} examination of X-ray flux from putative close 
binary systems;  and {\it vi)} RV measurements. In addition, we took advantage of 
CoRoT's capability to provide  colour information on transit events.

\subsection{Time-series photometric followup} 

The  \corot\,  exoplanet channel has a large \emph{PSF}. In the case of the \corots\, target, the 
FWHM is 8.6 arcsec along the dispersion axis and 6.0 arcsec  perpendicular to it.  Ninety-nine percent of the flux 
is extended over a larger and roughly ellispoidal area of 60 arcsec $\times$ 32 arcsec (Fig. 
\ref{Fig_CFHTadd} - left). This large area implies a significant probability that candidates detected 
in the \corot\, data arise from nearby background eclipsing binaries (BEBs). A photometric follow-
up programme of  \corot\, candidates intends to identify such BEBs, comparing observations during 
predicted transit-times with observations out of transit. This follow-up programme, as well as the time-
series follow-up performed on  \corotseven, are described in more detail in \citet{2009arXiv0907.2653D}. Here we only give a summary.

For any catalogued nearby star around the  \corotseven\,  target, we calculated the expected 
eclipse amplitude \emph{ if} this star was the source of the observed dips. Calculation of this 
amplitude is based on a model of the stellar \emph{PSF}, the shape of the photometric aperture, 
and the position and magnitudes of the target and the contaminating stars, respectively.  Several 
stellar candidates for false alarms were identified that way, with expected eclipse amplitudes 
between 0.2 and 1 mag.

Observations to identify such alarms were then done on two telescopes:  IAC-80 and CFHT. 
Images with the IAC-80 were taken on several occasions of the observed periodic loss of flux 
between February and April 2008; CFHT observations with \textsc{MEGACAM} 
\citep{2003SPIE.4841...72B} were performed during the ingress of a transit on 7 March 2008. From both data sets, 
photometric LCs  were extracted through classical differential photometric 
techniques and the stars on- and off-transit brightness are compared.
The observations from IAC-80 showed that none of the known contaminating stars could have 
been the source of an alarm, with all of them varying several times less than the amount required 
for a false positive. The CFHT time-series images (Fig~\ref{Fig_CFHTadd}) also shows a 
faint contaminator of about V = 19.5  some 10 arcsec   north of the target. However, this faint star 
would have to show strong variations of its brightness by a factor of 5 -- 8 to become a false alarm 
source, something that can clearly be excluded. Follow-up from photometric time-series imaging 
therefore allowed any false alarm to be excluded from sources
at distances over about 4 arcsec \ from the target.

\subsection{High angular resolution imaging followup}
  \begin{figure}
   \centering
\includegraphics[width=8.5cm]{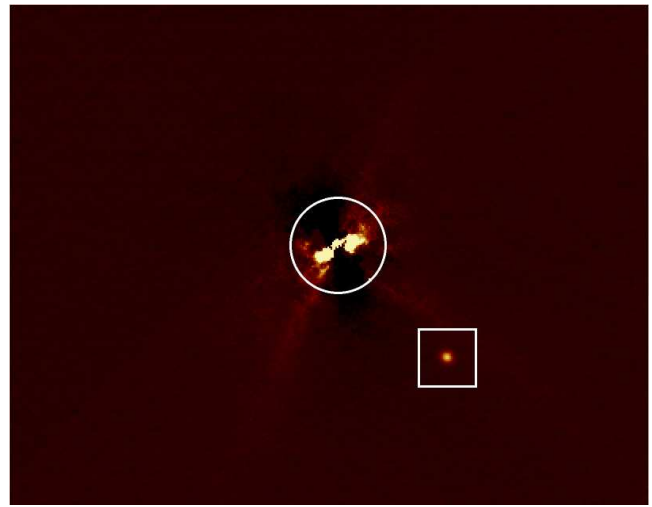}
      \caption{Central part of the NACO image,  after a 
      simulated  star 6.5 magnitudes fainter than \corotseven\,  was added (within square).  The 
circle of 400 mas radius defines the only region in which such a star could be confused with 
residual speckles.  
      \label{fake_400mas}
      }
   \end{figure}

The next step is to search for additional faint stars closer to the target   that might 
be potential sources for false alarms. This test employs high-resolution imaging with three 
different kinds of
observations: construction of the best image from the CFHT set,  sharp short exposures images 
taken with 
FASTCAM at the 1.5m CST and finally adaptive optics imaging with NACO at VLT.

In the first case we made a sub-pixel recentring  of all 100 MEGACAM images and took the 
median image. The result is shown in Fig. \ref{Fig_CFHTadd}. Two very faint stars, invisible in 
single images, become apparent  at angular distances of 4.5 and 5.5 arcsec from \corotseven\,  
(indicated with arrows in Fig. \ref{Fig_CFHTadd}). 
By adding simulated stars with known brightnesses at similar angular distances, we estimated 
them to be about 10 mag fainter than \corotseven, which is too faint to be potential alarms, even if 
they were to totally vanish during the transit.

FASTCAM is a lucky imaging camera \citep{2008SPIE.7014E.137O}. Here we only report on the 
deeper observations taken at the NOT, where 12,000 images, each with 50\,ms exposure time, were 
obtained on 24 October 2008 in I band, with a pixel resolution of 32 mas. The best result was
obtained from a  selection of the best 10\% of images followed by their centring and co-adding. 
Based on the absence of signals with an S/N higher than 5, the presence of relatively bright nearby 
objects with  I $\le$ 15 could be excluded beyond 0.18 arcsec, I $\le$ 16 beyond 0.3 arcsec
and  I $\le$ 17 beyond 0.8 arcsec. 
However, significantly fainter objects would not have been detected at any larger distance.

The VLT/NACO observations were performed thanks to discretionary time granted  by ESO (DDT 
282.C-5015).  A  set of J-band images with a pixel size of 13 mas was taken at different angles of 
the NACO rotator 
(15$^{\rm o}$ step), in jitter mode. The images are  recentred at sub-pixel level and median-
stacked for each rotator angle. The median of these stacked images gives essentially the 
\emph{PSF} of  
\corotseven, as all other objects in the field are removed by the median operation. Each stacked 
image is then substracted from this \emph{PSF}  and is de-rotated before a final median-stacked 
image is produced. This resulting image is shown in Fig. \ref{Fig2}: it reveals 3 faint stars (circled) 
at 
distances of 4.9, 5.9 and 6.7 arcsec,  whereas no other star could be indentified between 0.5 and 
5 arcsec. Clearly two of those stars are the counterpart of the two stars detected on the CFHT 
stacked image: the small difference in astrometry can be explained by our using of the 
average pixel size of Megacam, which is not constant throughout its very wide field of view. 
Photometry -- cross-checked against added simulated  stars -- shows that the  brightness 
difference between these stars and  
\corots\, is  $\Delta m_{\rm J}$ = 8.1 --  8.4.  
If we take m$_{\rm J}$(\corotseven) = 10.3, this translates to m$_{\rm J}$(faint stars) = 18.4 -- 18.7. 
On average, stars in the neighbourhood  exhibit V-J = 1.7, so that, if we apply this reddening to 
the three faint stars we find :  m$_{\rm V}$(faint stars) = 20.1 -- 20.4. Indeed the true reddening of 
those  three very faint stars is likely larger: either those stars are nearby low mass stars  and thus 
are fairly red, or they are very distant and thus suffer a large interstellar extinction. Their brightness 
difference to \corots\, in V of  $\Delta m_{\rm V}$ $\approx$ 10,  is thus consistent  with what is 
found  on 
the CFHT stacked  image. Even if these stars undergo a 50 \% decrease in brightness, they 
would produce only an amplitude variation  $\Delta$F/F = 5 10$^{-5}$ in the \corot\ LC, i.e. much 
less than the observed value of 3.35 10$^{-4}$.
We conclude that  none of the detected sources in the field around  \corotseven could account for 
the dips in the \corot\, LC, even if it would vanish totally. 

We now have to assess the probability of a contaminator still closer to the target. This case would 
correspond to a background of foreground system of a star and a transiting object (planet or star), 
the star having the same colours as \corotseven.   For instance it could be a star instrinsically bluer 
than \corots\,  but distant enough to be 
both reddened and faint enough to provide the observed signal.  In that case, the star 
should be at maximum  6.5 magnitudes fainter than \corotseven\, in J, taking the reddening into account 
and assuming that  its flux could be reduced by 50\% at maximum  to mimic a 
transit (case of a fully symmetric EB). To assess this case, we added on the NACO image   a 
simulated  star 
6.5 magnitudes fainter than 
\corotseven\, as shown in Fig. \ref{fake_400mas}. The simulated  star shows up 
clearly, brighter than any residual speckles farther than 400 mas from \corotseven. 
We can conclude that if it is a background binary system that mimics the observed transit on 
\corotseven\,, it   must be inside a circle of 400 mas radius, because at any other location it would have been 
seen. The probability  p that this is the case is simply the ratio of the surface of the 400 mas radius 
circle to the surface of the \corot\, mask: p = $\pi$ 0.4$^{2}$ / 640 = 8 10$^{-4}$.  The additional 
condition of similar colours for the  \corotseven\, and the contaminant makes this probability even 
lower, but, conservatively, we keep the preceding value\footnote{An independent estimate of this 
probability of a false positive can be made using CoRoTLux 2007 \citep{2007A&A...475..729F}. With 
that model, we computed the probability that a background eclipsing object is located at a distance 
0.4 arcsec from a given CoRoT target, with an amplitude that produces an apparent transit depth 
lower than 5 10$^{-4}$ and with an SNR above the CoRoT detection threshold (defined in  
\citealp{2009arXiv0903.1829A}, this volume). We obtain an average of 4 10$^{-4}$ object in the 
simulation of 
the LRa01 field ($\approx$10$^{4}$ stars) that exhibits such a small and detectable transit, a 
probability compatible with the upper limit we find here.}.

\section{ \corot\, colours 
}
\label{sec:corot_colors}

  \begin{figure}
   \centering
\includegraphics[width=8cm]{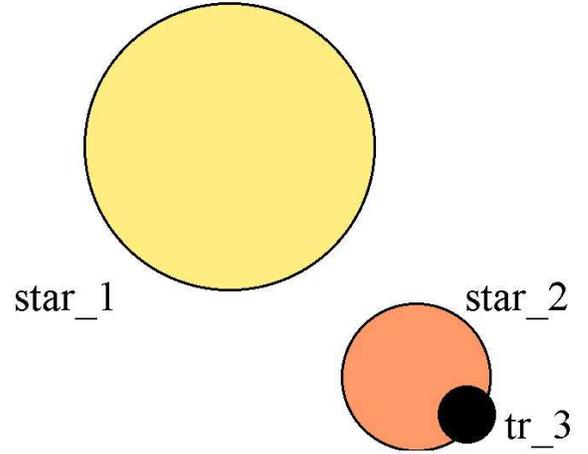}
      \caption{ Scheme of a triple system that could mimic the transit of a small planet in front of the 
      target    star (star\_1). Star\_2 is a physically associated faint star and tr\_3 a dark transiting 
      object, e.g. a hot Jupiter   or a brown dwarf.     \label{triple}
      }
   \end{figure}
   
  \begin{figure}
   \centering
\includegraphics[width=8.5cm]{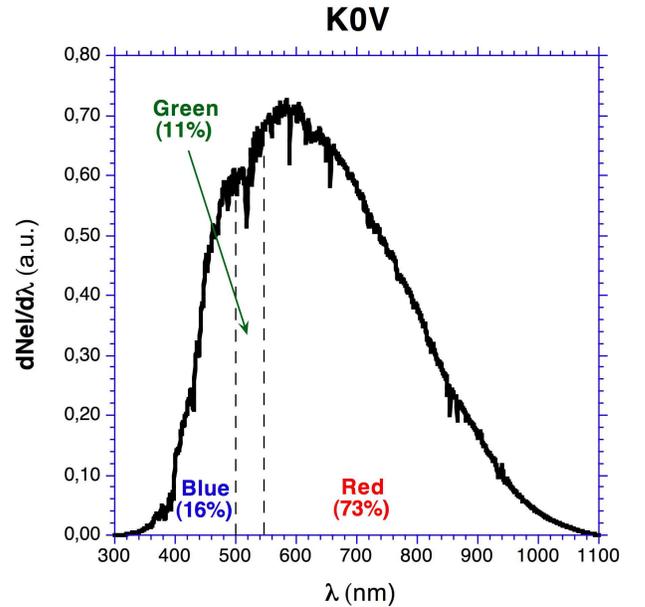}
      \caption{  Distribution of the flux of a K0V star (proxy for \corotseven\,, a G9V star) into the 3 
channels, according to the
       measured relative intensities of the coloured fluxes in photo-electrons.    \label{color_K0V}}
   \end{figure}

  \begin{figure}
   \centering
\includegraphics[width=8.5cm]{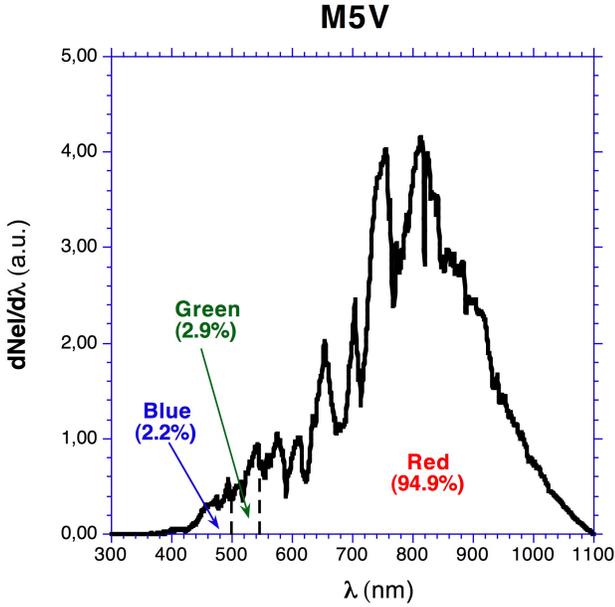}
      \caption{  Expected distribution into the 3 colours of the flux drop, $\Delta$F, if it was due to the 
eclipse of a M5V star. Frontiers between colours are the same as in Fig. \ref{color_K0V},  but the 
stellar spectrum is different. In that triple system, the flux drop would be significantly redder than 
observed.  
          \label{color_M5V}}
   \end{figure}

 \begin{figure}
   \centering
\includegraphics[width=8.5cm]{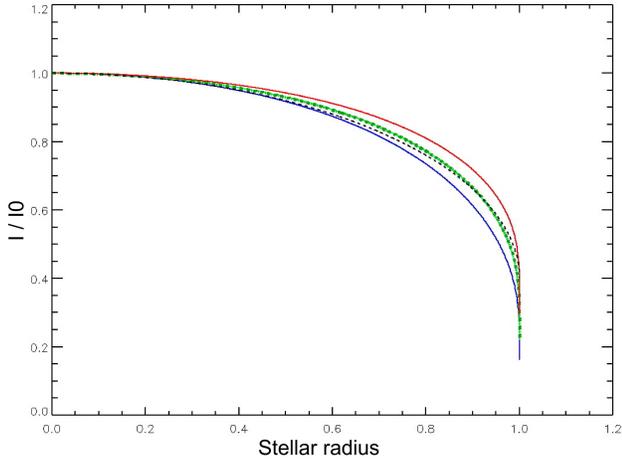}
      \caption{  Limb darkening effect versus fractional radius of the star: superimposition of the 
Eddington's classical law (black dash line) and of the predicted ones for a K0 star (best proxy in 
Claret's tables for a G9V star) in filter V 
(blue line), R (green line),  I  (red line) and their mean value (thick green, in fact  almost coinciding 
with R). The data for the predicted laws are from \cite{2000A&A...363.1081C}, using \teff = 5250 K, 
log g = 4.5, M/H = 0.0.       \label{Limb_drk_claret_Eddington}
      }
   \end{figure}

While the probability that there is a background star closer than 400 mas is low ($< 8 10^{-4}$), 
the probability that \corotseven\, is a triple system is significantly higher. A detailed study of stars 
by \cite{2008MNRAS.389..925T} shows that at least 8\% of the solar-type stars have three or more 
components. 
 
Illustrated in Fig. 9, the triple system could be a giant planet or a brown dwarf (tr\_3) transiting 
in front of a fainter star (star\_2) physically associated with the target star (star\_1). Assuming that 
the transiting object has a Jupiter size radius, the spectral type of the secondary star can be 
estimated thanks to the required brightness differences between (star\_1) and (star\_2). 

Using the Eddington's approximation for the  limb-darkening effect (Fig. 
\ref{Limb_drk_claret_Eddington}),  
the maximum reduction of star\_2 flux is \\
$(\Delta$F$_{2}$/F)$_{max} =  1.25 [0.4 + 0.6(1-z^2)^{1/2}] (R_{3} / R_{2})^2$, \\
where $z$ is the impact parameter of the transit on star\_2.   Assuming a mean value for $z$ = 0.5 
(not necessarily that estimated in section \ref{sec:planet_param}), the maximum flux reduction as 
measured by \corot\, is\\
$(\Delta \F / \F)_{max} = 1.15 (\R_{3} / \R_{2}(\M))^2 \F_{2}(\M) / [\F_{1} + \F_{2}(\M)]$, \\
where the radius and flux of star\_2 are a function of its mass M.
Assuming that the transiting object, either a hot Jupiter or a brown dwarf, has a Jupiter radius, $\R_{3}$ = \RJ,  
and  using the radius and the flux of a mean sequence star \citep{2000asqu.book..381D} for star
\_2, and the  \corot\, 
measured  $(\Delta \F / \F)_{max} = 3.5\, 10^{-4}$, the preceding relation can be solved as an 
equation in M. 
The  found star spectral type is M5.1V, approximated as M5V. The assumption that the 
transiting object has a Neptune size would lead star\_2 to be a K9V star. Now, such a star would 
be redder than the target star\_1, providing a criterion to qualify / falsify the hypothesis.

The details of the argumentation are explained in a dedicated paper (Bord\'e et al, in preparation). 
Only the principles and the results are reported here. The bi-prism of \corot\,   
produces a mini-spectrum for each star, split by a proper selection of pixels  into 3 spectral 
bands whose fluxes are recorded independently \citep{2000ESASP.451..221R,
2006ESASP1306..283A}. The boundaries within the 
photometric mask that define these colours are chosen so that the red, green, and blue parts 
correspond, as much as possible, to given fractions of the total, but they must also correspond to 
an integer number of columns on the CCD (dispersion is done in rows). In the case of the G9V 
target star (star\_1)  that dominates the total flux, the actual fractions  are 73.3\%, 10.9\%, and 
15.8\%, 
respectively, as indicated by the number of photoelectrons in the different channels (Fig. 
\ref{color_K0V}).

Assuming that the M5V star\_2 is part of the triple system,  it would be at an angular position so 
close to the target star that it would be indistinguishable with the \corot\, spatial resolution. The 
same boundaries on the CCD for defining the colours would then apply. 
 This transiting star would lead to photoelectron contents in the 3 bands that are different from 
those of the target.
They can be estimated from the spectrum of an M5V star (Fig. \ref{color_M5V}). The red fraction 
would increase to 94.9\% and the green and blue decrease to  2.9\% and 2.2\%, respectively. 
These fractions correspond to the
expected  transit flux variations in that hypothesis, $(\Delta F)_{Red}$, $(\Delta F)_{Green}$, and $
(\Delta F)_{Blue}$.

If the r, g, b quantities are defined as\\
$r = \frac{ (\Delta F / F)_{Red}}{ (\Delta F / F)_{White} } $, $g = \frac{ (\Delta F / F)_{Green}}{ (\Delta 
F / 
F)_{White} } $ and $b = \frac{ (\Delta F / F)_{Blue}}{ (\Delta F / F)_{White}}$, \\
their expected and observed values can be compared. To work with a sufficient S/N, 
we bin 150 colour transit curves into 15 bins of 10 LCs each, calculate the $r$, $g$ and $b$ 
values, make the corresponding histograms and estimate the observed mean values and 
standard 
deviations. The observed and expected values  are\\
$r_{obs} = 0.88 \pm 0.18,	r_{M5} = 1.29 ; \\
g_{obs} = 1.24 \pm 0.30,  	g_{M5} = 0.27 ; \\
b_{obs} = 1.42 \pm 0.41,	b_{M5} = 0.14.$ 

We conclude that the observed colours are incompatible with those of a transit in front of an M5V 
star, whereas they are compatible with a transit in front of the target star ($r_{\rm G9} = g_{\rm G9} = b_{\rm G9} = 1$)\footnote {If a circular orbit is assumed, the mere duration of the transit requires that the star undergoing the 
transit has a minimum size. The observed duration of the transit is a fraction  $f= 0.061$ 
of the period. $f$ also satisfies  $f \le (R_{2}+R_{3}) / \pi a$, the equality corresponding to an equatorial 
transit. For a Jupiter-size transiting object, or smaller, this translates into R$_{2} > 0.72$ R$_{\odot}$, 
corresponding to a star\_2 earlier than K6V, if it is a main sequence star, a statement that is stronger than 
the one we derive but that needs the additional assumption of a circular orbit}.
Another possibility could be that star\_2 is a white dwarf with, by chance, the very temperature of 
the target star. With a proper luminosity ratio, this could be indistinguishable from a small planet in 
front of the target star,  just from the colour criterion. However, this situation would produce a 
transit duration, 1 mn, much shorter than what is observed, because a white dwarf has a similar size to  a terrestrial planet, not that of a main sequence star. The white dwarf possibility is then 
discarded without ambiguity.
 \textit{We conclude that  the observed LC 
is not produced by a triple system where a Jupiter size object transits in front of a secondary star}.

\section{Infrared spectroscopy 
}
\label{sec:spectro}

The reasoning of Sect. \ref{sec:corot_colors} can even tell more about the possible spectral types 
of star\_2. The observed colours of the transit permit to restrict these to earlier types than M0V 
because later stars would produce ratios g $<$ 0.55 and b $<$ 0.45  that are not compatible with 
the observations (mean values $\pm 2 \sigma$). 

We can now conclude that the remaining possible star\_2 that could still produce a false 
positive is rather bright. This could produce observable signatures in the target spectrum, 
particularly in the IR where the contrast between a G9V and M or K type star is lower, so we 
took these spectra of the \corotseven\, star.

Thanks to ESO DDT time (DDT 282.C-5015), we obtained a high-resolution spectrum of 
\corotseven\, using the infrared spectrograph CRIRES mounted on the VLT-UT1 (Antu).
The AO-system was used along with a slit-width of 0.3 arcsec that resulted in spectral resolution of 
R = 60,000. The wavelength coverage was 2256 to 2303 nm, a region that included the CO-
overtone lines, which have an equivalent width about a factor of three larger in an M5V star than in 
a G9V star. The total integration time of 1800 s resulted in an S/N = 100 per resolution element. 
Standard IRAF routines were used for flat-fielding, sky subtraction, and wavelength calibration. A 
spectrum of HD48497 taken with the same setup during the same night and at similar air mass 
was used  to remove the telluric absorption lines.

\begin{figure}[h]
   \centering
\includegraphics[width=8.5cm]{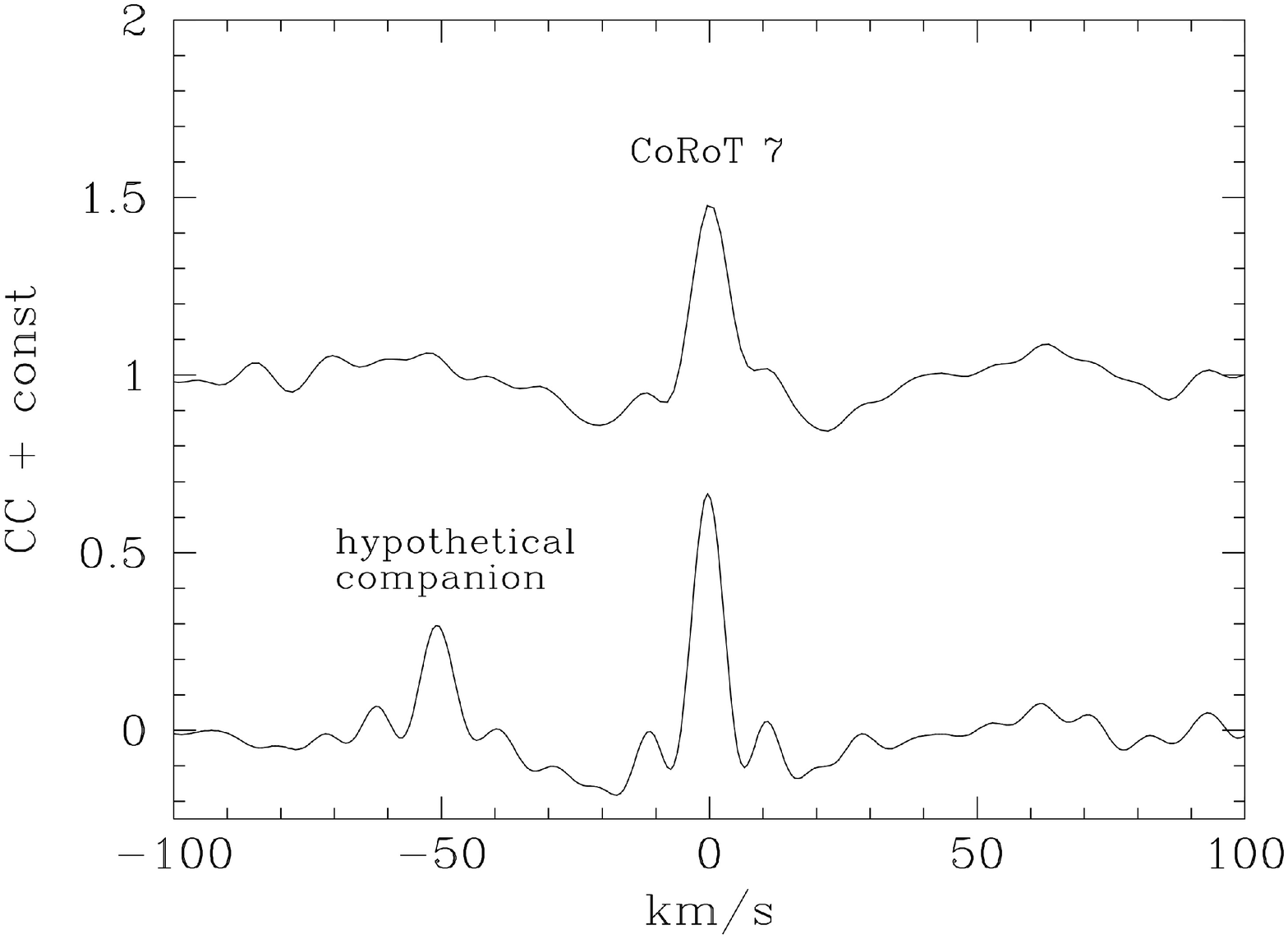}
\caption{ Top: cross correlation function between the 
CRIRES spectrum of \corotseven\, and a synthetic spectrum of an 
 M5V star. Bottom:  same as above, but an  M5V star spectrum  with a shift in RV of -50 km/s was 
 added to the \corotseven\, spectrum. We conclude that there is no such companion.
\label{crires}}
\end{figure}

Figure \ref{crires} (top) shows the cross-correlation function of the \corotseven\, spectrum with the 
theoretical spectrum of an M5V star. Since  we could not find an M5V template in the ESO archives, we simply  calculated the theoretical spectrum by using the spectrum of an M2V 
star and increased the strength of the CO lines to the strength of an M5V star.  If there is an early 
M star  associated to \corotseven, we would expect a secondary maximum exhibiting a difference 
in RV 
with respect to \corotseven. This difference would be less than 
about 100 km/s, since a system with a shorter orbital period would not be stable. 
Figure \ref{crires}  (bottom) shows  the cross-correlation function when a putative M5V
star at the same distance as \corotseven\, and with an RV difference  of 50 km/s  added. With this 
analysis, we conclude that, if the separation in RV of the two stars is greater than 8 km/s, the 
primary and secondary peaks would be well separated. If the separation was between 3 and 8 
km/s, we would still detect an asymmetry of the cross-correlation function. 

An independent analysis of 
the CRIRES spectrum using TODCOR \citep{1994ApJ...420..806Z}   was able to exclude the 
presence of an M star with a brightness as low as 7\% of \corotseven, thus more than a factor of 
two lower than needed to rule out the presence of a bounded M-star. Such a secondary M star is 
clearly detected by TODCOR analysis when its spectrum is inserted with a proper weight to a 
sunspot template, a proxy for an M-star.

Therefore, the CRIRES spectra can rule out the presence of a secondary star earlier than M5V 
within 0.3 arcsec (CRIRES slit width) from \corotseven\, except if, by chance, observations 
were performed when star\_1 and star\_2 had the same RV within 3 km/s. 
Assuming  that \corotseven\, is a binary consisting of a G-star  orbited 
by an M-star that has a eclipsing planet, then the probability that we observe the system at the 
very moment of the conjunction so that the separation is less than 3 km/s  is below 10$^{-6}$. 

Eventually, CoRoT colours allow us to exclude a secondary companion fainter (redder) than M0V, 
and CRIRES spectrum a companion brighter than M5V, but for a very special case whose 
probability  is less  than 
10$^{-6}$. As the exclusion intervals overlap, we can conclude that the observed events are not 
due 
to an associated star (star\_2) subjected to a transit. The triple system hypothesis is essentially 
rejected.

\section{Lack of X-ray emission }

The case of a triple system made of the main target and two grazing eclipsing binaries  is already 
practically excluded by Sects. \ref{sec:corot_colors} and \ref{sec:spectro} and the U shape of the 
transit. 
An additional hint is provided by the lack of X-ray emission. If  \corotseven\, were a triple system 
consisting of a G9V star and an eclipsing binary of late spectral type with an orbital period of $
\approx$ 0.9 
day, the binary would likely be detectable in the X-ray spectral domain. As a prototype object of 
this kind, 
we consider the eclipsing binary YY Gem consisting of two M1Ve stars with an orbital period of 
0.85 days  \citep{1990A&A...230..419H,2002A&A...392..585S}.  The brightness of YY Gem in the 
X-ray regime is 2 - 8 10$^{29}$ erg/s, and its distance is 15.8 $\pm$ 0.30 pc 
\citep{2002A&A...392..585S}.  
\citet{1993A&AS..100..173S} in their catalogue of 
chromospherically active binaries list at least 7 late-type stars (late G- to early M-type) with 
rotational (binary) periods of less than 1 day. All of these have X-ray luminosities in the range 0.3 
-- 0.9 $\times$ 10$^{30}$ erg/s. Using the data obtained in the ROSAT all sky survey 
\citep{1999A&A...349..389V} in the 0.1 to 2 keV band, we searched for possible X-ray emission 
from hypothetical companions of \corotseven .  
With the lowest X-ray luminosity of such short-period binaries, we would have detected these 
companions of \corotseven at 4 $\sigma$ level, if present. Using  d(\corotseven) = 150 
pc, we can state that $L_{X}$(\corotseven)$ < 5 10^{28}$ erg/s, that is the level of the sky 
background above which one cannot find any evidence of a source at the location of the target. 

\section{Radial velocity campaign}
\label{sec:RV}

\begin{table}[h]
\begin{center}{
\caption{SOPHIE RV observations of \corotseven.}
\begin{tabular}{llll}
\hline
Julian date & Orb. & RV & 1$\sigma$ error\\ 
day       & Phase & km/s &  km/s \\
\hline
2454514.32514  & 0.19 &  31.1444& 0.0095\\  
2454517.40230  & 0.79 &  31.1392& 0.0073 \\ 
\label{Tab-Sophie}
\end{tabular}}
\end{center}
\end{table}

\begin{figure}
\begin{center}{
\includegraphics[width=8.5cm]{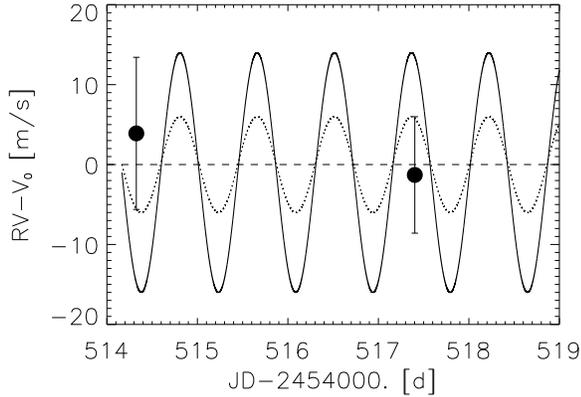}
\caption{SOPHIE RV measurements of CoRoT-7 obtained in February 2008, versus 
time. The two data points are shown with 1$\sigma$ error bars.  Superimposed are Keplerian 
orbital curves of a 0.854 day period planet at the CoRoT ephemeris, which are excluded by the 
data at 1$\sigma$ (dotted line) and 2$\sigma$ (plain line). They correspond to semi-amplitudes of  
6 and 15 m/s, or planetary masses of 8.5 and 21 \ME. The systemic velocity $V_0$ is left as a free 
parameter in this simple fit. We conclude that the planet, if present, has a mass \mplanet $<$ 21 
\ME\, at a 95 \% confidence level.
}
\label{RV-Sophie}}
\end{center}
\end{figure}

We have obtained two measurements of \corotseven\, with the spectrograph SOPHIE 
\citep{2006tafp.conf..319B} at the 1.93m telescope in Observatoire de Haute Provence (France). 
The RV observations were conducted on 17 and 20 February 2008, just a few weeks after the early 
discovery of the transiting candidate and during the CoRoT observing run.  RV follow-up observations were assigned high priority because of  the shallowness of the 
transit. 
The SOPHIE RV measurements are given in Table \ref{Tab-Sophie} and shown in 
Figure \ref{RV-Sophie}. The two measurements were obtained on CoRoT dates 514.3 and 517.4 
JD-2 454 000 (see Fig. \ref{Corot_LC}), i.e., at a time when the photometric activity from the star is 
relatively low; consequently, they are not strongly affected by this variability. Quantitatively, the 
photometric difference measured by CoRoT between the two SOPHIE observations is 0.2\%, 
which  is 10 times smaller than the largest variation in the total CoRoT light curve, and twice 
smaller than the mean standard deviation. The difference in RV between the two measurements is 
only 5 $\pm$ 10 m/s, although they were obtained at almost extreme phases (0.19 and 0.79, when 
the transit occurs at phase=0). When fitting the systemic velocity and radial-velocity semi-
amplitude of a circular orbit at the CoRoT ephemeris, \textit{we can exclude a 
planetary mass higher than 21 \ME\, at 2$\sigma$}. 

Due to the lack of detected RV signal at the level of 10 m/s (1 $\sigma$ ), it was decided to 
observe \corotseven\,  with HARPS, which has a higher velocity accuracy, at the next observing 
season starting October 2008. More than 100 measurements were taken and the results will be 
described in another paper (Queloz et al., in preparation).  

The following conclusions can be drawn from the RV SOPHIE measurements\footnote{It is 
interesting to note that \textit{we do not need this RV information} formally. The case of a grazing large 
object, e.g. a Jupiter or a late star, can be eliminated because the corresponding transit duration would
be significantly  shorter than observed. In the $\chi^2$ map resulting from the transit fit using  the 
orbit inclination and planetary radius as free parameters, we 
find a minimum at the values given in Sect. \ref{sec:planet_param} but \textit{no} secondary minima that would 
correspond to larger inclinations and radii. For instance, a grazing Jupiter would give a  22 min 
long transit -- assuming the limb darkening of Fig. \ref{Limb_drk_claret_Eddington} -- which is 
excluded by the \corot\, observations.  }:\\
(i) there is no Jupiter mass planet or, \textit{a fortiori}, a stellar companion, bound to CoRoT-7. A 
Jupiter mass object with a 0.854day period would produce a 230 m/s RV signal and a  white dwarf companion an amplitude over 100 km/s. 
None of these are observed; \\
(ii) the data are compatible with a planet with a 0.854day period and a mass of less than 
21 \ME, provided that no other change in the RV of the star occurs between these two 
observations (due to other companions or stellar activity). A formal detection and an accurate 
estimate of the mass is discussed in Queloz et al (in preparation). 
\\

As a result of Sects. 3 to 7, where we excluded almost all the cases of false positives, we conclude that a small planet  
orbits the star \corotseven\, with a 0.854 day period, with a risk of false positives conservatively 
estimated to be $<$ 8 10$^{-4}$.

\section{Stellar parameters}
\label{sec:stellar_params}

The central star was first spectroscopically observed in January 2008 with the 
AAOmega multi-object facility at the Anglo-Australian Observatory. By 
comparing the low-resolution ($R /approx 1300$) AAOmega spectrum of the 
target with a grid of stellar templates, as described in \citet{2003A&A...405..149F} 
and \citet{2008ApJ...687.1303G}, we derived the spectral type and luminosity class 
of the star (G9 V). These observations allowed us to asses the dwarf nature 
of the target, rule out the false positive scenario of a low-mass star 
orbiting a giant star, and trigger further and systematic high-resolution 
spectroscopic follow-up of the system.

A preliminary photospheric analysis of the central star was carried out using a \uves\ spectra 
registered on October 2008 (Programme 081.C-0413(C)). The resolving power of this observation is  
$\simeq$ 75,000 and the S/N is about 100 per resolution element at 
5500\AA. 
Later, we also took advantage of a series of 80 \harps\ spectra acquired during the RV 
monitoring of the target.
 Even though the detailed analysis of those more recent specta is still in progress, the first results we 
achieved are in good agreement with those derived from the UVES spectrum analyses and are 
also presented in this paper.
 \\
 \\
\textit{Abundances}

Using the VWA method \citep[see][and references therein]{bruntt08} a preliminary abundance 
analysis was carried out from non-blended lines. The derived abundances calculated relative to 
the solar ones are given in Table~\ref{VWA0165}. 
They indicate [M/H] = + 0.03 $\pm$ 0.06, i.e.  solar-like. This spectral analysis also shows that 
\corotseven\,   is a slowly rotating main-sequence star late G main-sequence star with nearly solar-
like abundances. 
\\
\\
\textit{Mass, radius, and effective temperature}

To determine the atmospheric parameters of the star, we used the same approaches as 
for the other \corot\ host stars \citep[see][e.g.]{2008A&A...491..889D}, with different groups carrying 
independent analyses using different methods. 
The mass and radius of the star were determined from the photospheric parameters derived from 
our spectroscopic analysis combined with evolutionary tracks in the H-R diagram. 

Recent studies have clearly demonstrated that the luminosity  in transiting systems can be very 
well constrained by the LC  fitting \citep{Pont2007,Sozzetti2007}. However, for \corotseven\,   
the shallow transit and the stellar activity result in a large uncertainty on the stellar density 
and further on the planet radius, so we have abandoned it. 

On the other hand, the {Na \sc i} D and {Mg \sc  i} line wings in the spectra yield good 
constraints on the measured \logg\ value, $\pm$ 0.10, an accuracy already obtained by other 
authors, e.g. \citet{Sozzetti2007}, so we used our spectroscopic estimate of the surface gravity 
(\logg\  = 4.50)  as a proxy for the luminosity. 

The grid of the STAREVOL stellar evolution models  \citep{2006A&A...448..717S} was interpolated 
within the locus defined by the three basic 
parameters (\teff, \mstar, \rstar) and their associated errors. The resulting stellar parameters are 
reported in Table~\ref{StarParam}.

\begin{table}
 \centering
 \caption{Abundances of 21 elements in Corot-Exo7,  from VWA. Last column indicates the 
number of lines.
 \label{VWA0165}}
 \setlength{\tabcolsep}{3pt} 
 \begin{footnotesize}
\begin{tabular}{l|lr|lr}
\hline
  {C  \sc   i} &    $+0.11  \pm0.36$  &   3  \\ 
  {Na \sc   i} &    $+0.02  \pm0.08$  &   4  \\ 
  {Mg \sc   i} &    $+0.07         $  &   1  \\ 
  {Al \sc   i} &    $+0.12         $  &   2  \\ 
  {Si \sc   i} &    $+0.05  \pm0.04$  &  21  \\ 
  {Ca \sc   i} &    $+0.09  \pm0.05$  &   7  \\ 
  {Sc \sc   i} &    $-0.00         $  &   2  \\ 
  {Sc \sc  ii} &    $+0.03  \pm0.05$  &   5  \\ 
  {Ti \sc   i} &    $+0.06  \pm0.04$  &  36  \\ 
  {Ti \sc  ii} &    $+0.00  \pm0.05$  &  11  \\ 
  {V  \sc   i} &    $+0.17  \pm0.05$  &  16  \\ 
  {Cr \sc   i} &    $+0.04  \pm0.04$  &  23  \\ 
  {Cr \sc  ii} &    $-0.01  \pm0.04$  &   3  \\ 
  {Mn \sc   i} &    $-0.03  \pm0.05$  &  10  \\ 
  {Fe \sc   i} &    $+0.05  \pm0.04$  & 250  \\ 
  {Fe \sc  ii} &    $+0.04  \pm0.05$  &  18  \\ 
  {Co \sc   i} &    $+0.04  \pm0.05$  &  15  \\ 
  {Ni \sc   i} &    $+0.04  \pm0.04$  &  62  \\ 
  {Cu \sc   i} &    $-0.02         $  &   1  \\ 
  {Zn \sc   i} &    $+0.01         $  &   1  \\ 
  {Sr \sc   i} &    $+0.14         $  &   1  \\ 
  {Y  \sc  ii} &    $+0.09  \pm0.07$  &   3  \\ 
  {Zr \sc   i} &    $+0.01         $  &   1  \\ 
  {Ce \sc  ii} &    $+0.22         $  &   1  \\ 
  {Nd \sc  ii} &    $+0.04  \pm0.07$  &   2  \\ 
\end{tabular}
\end{footnotesize}
\end{table}

\begin{table}[h]
\begin{center}{
\caption{\label{StarParam} 
\corotseven\,   parameters derived from RV and spectroscopic analyses.}
\begin{tabular}{lll }\hline
\hline
\vrad\ (\kms)  & $+$31.174  & $\pm$ 0.0086 \\
$v_{\rm rot} \sin i$ (\kms) &  $<$ 3.5      &  \\  
\teff\ & 5275K & $\pm$ 75 \\
\logg\ & 4.50 & $\pm$ 0.10 \\
\met\ &  $+$0.03 & $\pm$ 0.06 \\
Spectral Type & G9 V\\
$M_\star$ & 0.93 &  $\pm$ 0.03 \\
$R_\star$ &  0.87  & $\pm$ 0.04 \\
$M_V$ &  5.78  & $\pm$ 0.10 \\
Age  & 1.2 - 2.3 Gyr &   \\
Distance &  150 pc &  $\pm$ 20 \\
\hline
\end{tabular}}
\end{center}
\end{table}

We also use the available visible and near infrared photometric data to estimate the effective 
temperature independently.
The map of neutral hydrogen  column density $N_H$ in the galactic plane
of \cite{1984AJ.....89.1022P} indicates a maximum value of $N_H  = 10^{20}$ cm$^{-2}$ for the line of sight of \corotseven\, within 200 pc, which corresponds to $E(B\!-\!V)\approx
0.01-0.02$ mag. The presence of a small amount of extinction in
this direction
is also confirmed by the maps of \cite{2003A&A...411..447L}.
Using Exodat  \citep{2009AJ....138..649D} and 2MASS photometry, the \cite{2006A&A...450..735M}
calibration yields \teff\,$ = 5300\pm 70$~K.
This effective temperature from broad-band colours  therefore agrees with
the spectroscopic determination reported in Table \ref{StarParam}.
\\
\\
\textit{Distance}

For the distance estimate we have first converted the 2MASS magnitudes into
the SAAO system with the relations of \cite{2001AJ....121.2851C}.
The calculated colours J-H and H-K are 0.474 $\pm$ 0.030 and
0.046 $\pm$ 0.029, respectively.
These colours are compatible with main sequence stars of spectral types
between G8 and K2.
Given the constraint on the spectroscopic measurement of \teff\,  our best estimate
of the spectral type is G9. Assuming for this spectral type an
absolute magnitude of
$M_V=5.8 \pm 0.1$ \citep{1981Ap&SS..80..353S} and extinction as already reported, we obtain  
an estimation of the
distance $d=150 \pm 20$ pc. 
\\
\\
\textit{Projected rotational velocity}

The projected rotational velocity is determined by fitting several isolated lines in the HARPS 
spectrum with synthetic profiles. The synthetic spectra are convolved by the instrumental profile 
approximated by a Gaussian function (R = 115,000), and a broadening profile comprised of 
macroturbulence and rotation. Since macroturbulence and rotation are strongly coupled, the value 
of  \vsini\,  is somewhat uncertain. We therefore explored a grid of values for macroturbulence. The 
possible range for the macroturbulence is 0 -- 3 km/s since higher values provide poor fits for all 
lines. For this range of increasing macrotubulence, the best fit of  \vsini\, values decreases from 3.5 
to 0 km/s. Our estimate is then \vsini\, $<$ 3.5  km/s.
\\
\\
\textit{Age}

The age estimate derived from the H-R diagram is poorly constrained. To overcome this limitation, 
we use different age indicators: Li {\sc i} abundance, the Ca {\sc ii} H and K chromospheric emissions 
\citep{Noyes1984}, and gyrochronology \citep{2007ApJ...669.1167B}.

In the \corotseven\, spectra, no Li {\sc i} line is detected (Fig. \ref{LiII}), even in the co-added 53 
HARPS exposures. This non-detection points to an older age than the 0.6 Gyr of the Hyades 
\citep{2005A&A...442..615S}.

\begin{figure}
   \centering
    \includegraphics[width=8.5cm]{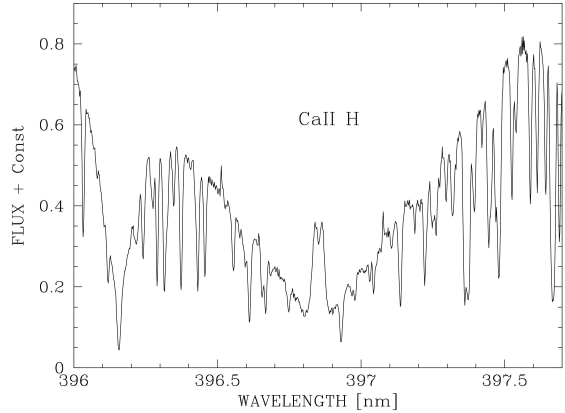}
   \caption{\caii\ H line emission as observed in the co-added \harps\ spectra of \corots\, .}              
\label{CaII}
    \end{figure}

\begin{figure}
   \centering
   \includegraphics[width=8.5cm]{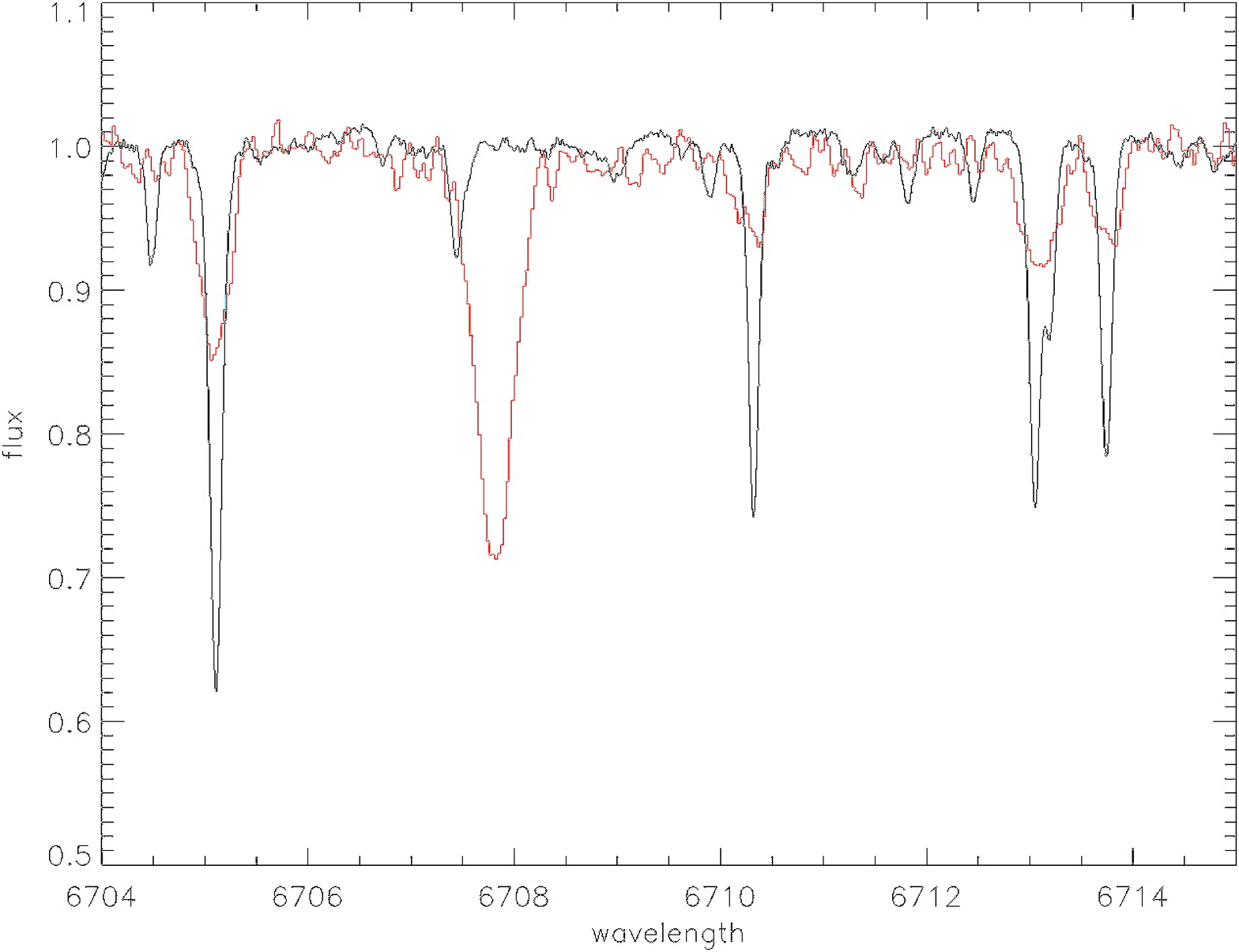}
   \caption{Comparison of \corotseven\,   (black line, co-added HARPS spectra) and \corottwo\,  spectra (red line, UVES spectrum) in a spectral window centered on the 
   \lii\ $\lambda$6707.8 doublet. While the \corottwo\, spectra displays a strong \lii\, feature, only the 
nearby \fei\,  line at
    6707.44\AA\, is visible in the co-added \harps\,  spectra of our target.
    }              
\label{LiII}
    \end{figure}

The activity of \corotseven\, is apparent not only in  the CoRoT light curve (Fig.\ref{Corot_LC}), but 
also in  the broad photospheric Ca {\sc ii} H \& K absorption lines (Fig. \ref{CaII}), which vary with 
time. For each HARPS spectrum, following the prescription of \cite{Santos2000}, we calculated the 
usual chromospheric flux index, log R$_{\rm HK}$, which measures the Ca {\sc ii} H \& K fluxes, 
converted to the Mount Wilson system, and corrected for the photospheric flux. Over the one-year 
period of our series of HARPS spectra, we estimated the mean stellar activity level to be log R
$_{\rm HK}$ = 
- 4.601 $\pm$ 0.05, with an uncertainty estimated from the range of observed values. Using the 
relations in \cite{Wright2004}, we derive a chromospheric age estimate of 1.4 $\pm$ 0.40 Gyr and 
a rotational period of 23 $\pm$ 3 days. We compared this chromospheric age estimate with the 
new activity-age relation given by \cite{Mamajek2008} that yields an age of 2.0 $\pm$ 0.3 Gyr. 
Both values are consistent within the error bars.

A fourth age estimate could be done from the stellar rotation rate. The rotation-age relation is often 
presented as of limited interest; however, \citet{2007ApJ...669.1167B} recently revised the method 
and proposes a procedure, called gyrochronology, which provides the age of a star as a function 
of its rotation period and colour. Using his formalism, we infer an age of 1.7 $\pm$ 0.3 Gyr using  
the rotational period of 23 days derived from the LC.
These different diagnostics all agree. We thus adopt an age between 1.2 and 2.3 Gyr. 
The measured chromospheric activity, higher than the solar value, also points to the idea
 that \corotseven\, is not an old, quiet star.
 
\section{Planetary parameters }
\label{sec:planet_param}

\begin{figure}[!th]
\begin{center}
\includegraphics[width=8.5cm]{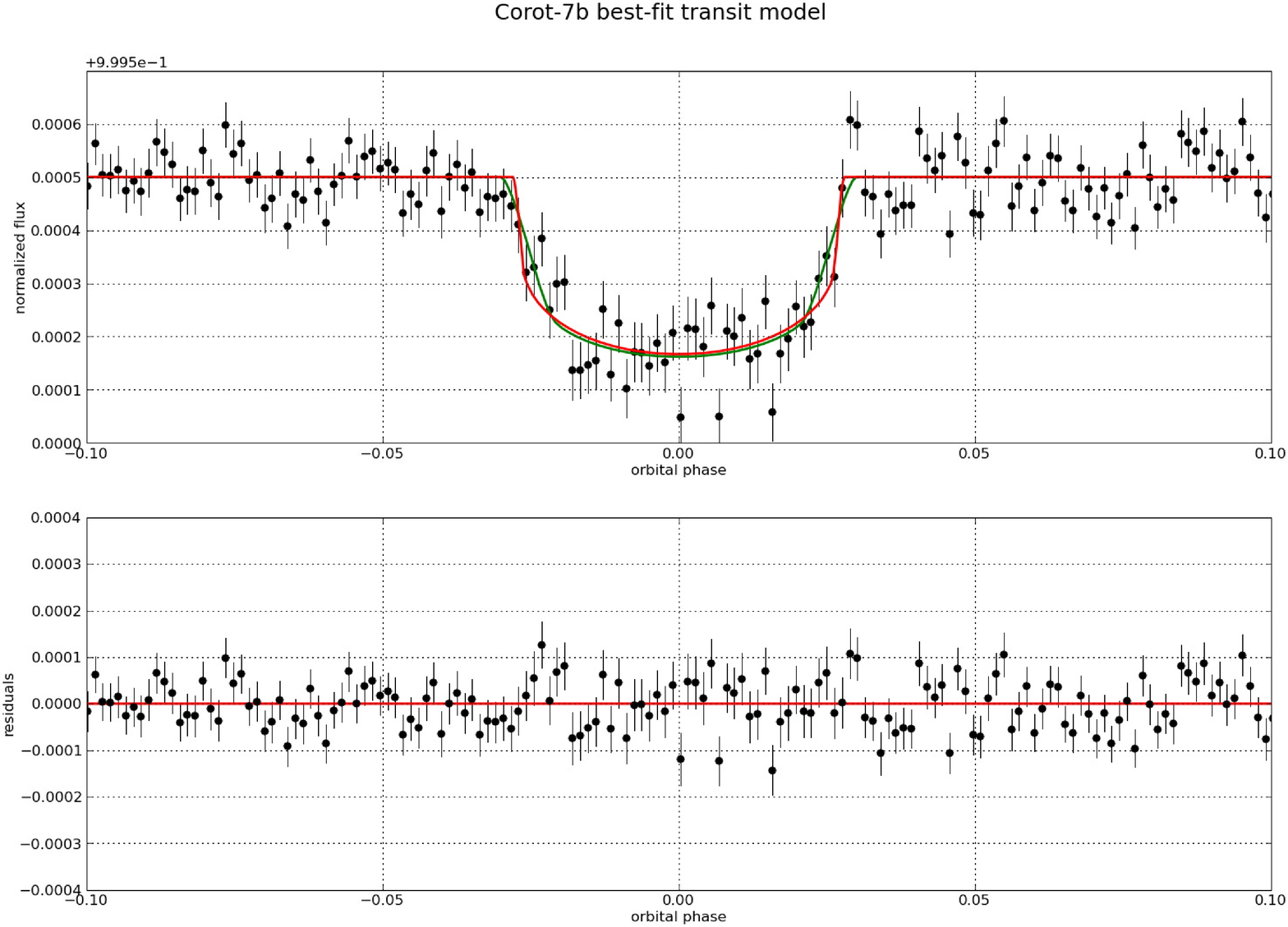}
\caption{Phase-folded LC  of \corotsevenb\, using the ephemeris given in Table~\ref{Tab_transit_param}, 
and combined in bins of $\sim$1.5~min. A fixed period (Table \ref{table_planet}) has been used. 
The green line is the 4-parameter best-fit model, using \cite{2006A&A...450.1231G}, but it leads to 
a stellar density in conflict with the one determined by spectroscopy. 
The red line corresponds to the finally adopted solution, leaving only inclination and planet radius 
as free parameters; the bottom panel shows the residuals of the fit. See the text for details.} 
\label{Corot_exo_0165_phasefolded_17feb09_v2}
\end{center}
\end{figure}

\begin{figure}[!th]
\begin{center}
\includegraphics[width=8.5cm]{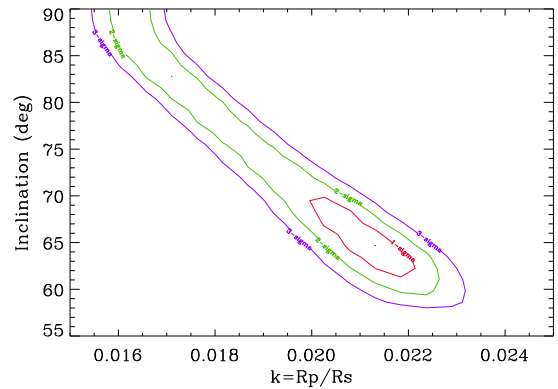}
\caption{Contour of the chi2 of residuals in the  space [relative planet radius, inclination], when a 
classical fit with 4  free parameters is used, fixing the stellar mass and radius at the spectroscopically
determined values. Contours at 68\%, 95\%, and 99.7 \% confidence level for parameter estimates
are plotted (red, green, and purple contours). 
One notes the high degeneracy of the secure solutions (green and purple contours).  
} 
\label{chi2-4paras}
\end{center}
\end{figure}

\begin{figure}[!th]
\begin{center}
\includegraphics[width=8.5cm]{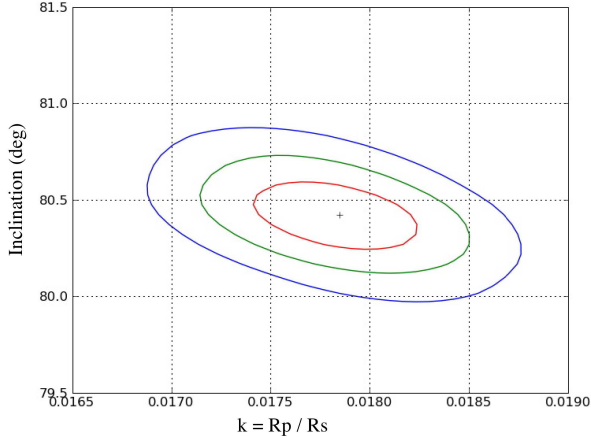}
\caption{Contour of the chi2 of residuals in the two-parameter space [relative planet radius, 
inclination], fixing the stellar mass and radius at the spectroscopically determined values. Contours at 68\%, 95\% and 99.7 \% confidence level are plotted.  There is no more 
degeneracy, so mean values and uncertainties of the fit can be derived.  The uncertainties 
on the stellar parameters are not taken into account here.   } 
\label{chi2-contour}
\end{center}
\end{figure}

We derive the planetary parameters using the stellar parameters of the previous section and the 
information derived from the  \corot\, \,LC. 
\\\\
{\it Semi-major axis}

Applying Newton's law and using estimates of \mstar and \rstar from Sect. \ref{sec:stellar_params}, 
we find \\
a  = $[\mmstar (\M_{\odot}) P(yr)^{2}]^{1/3}$ = 0.0172 $\pm$ 0.0002 AU,  a / \rstar = 4.27 $\pm$ 
0.20.\\
The uncertainties are mainly caused by the stellar mass and radius uncertainties, because the period is 
known to a high degree of accuracy.
\\
{\it Radius of the planet and inclination of the orbit}. 

Following the technique used for the other \corot\, discovered planets 
\citep{2008A&A...482L..17B}, all the observed transits 
are combined after a low-order polynomial (order 2 in this case) is fitted in the parts surrounding 
each transit and substracted. 
The period is fine-tuned by choosing the one that provides the shortest duration of the phase-
folded transit. The individual 
measurements are combined in bins of 0.0012 in phase, corresponding to about 1.5 min, and the 
error assigned to each 
binned point is estimated as the standard deviation of the points inside each bin, divided by the 
square root of the number of 
points .

We first used the formalism of \cite{2006A&A...450.1231G}, combined with the AMOEBA 
minimization algorithm \citep{press} in order to obtain a first evaluation of the transit parameters. 
Because of the moderate S/N of the curve, we did not try to fit for the quadratic limb-darkening 
coefficients, but instead we fixed them to values corresponding to a G9V star (from 
\citealt{2000A&A...363.1081C}, with 
$u_+ = u_a+u_b = 0.6$, $u_- = u_a-u_b = 0.2$). 
The four fitted variables were the centre of the transit, the ratio k=\rplanet\,  / \rstar, the orbital 
inclination $i$, and the phase of transit ingress $\theta_1$, which can be translated into the 
scale of the system a/\rstar using Eq.~12 of \cite{2006A&A...450.1231G}.
Figure \ref{Corot_exo_0165_phasefolded_17feb09_v2} shows a fit of the transit as the thin 
green line.  Under the assumption of a circular orbit,  the scale of the 
system,  a/\rstar=1.9$\pm$ 0.1, can be translated into a density of 0.17 g cm$^{-3}$ for the host 
star, a value much lower than expected for  a G9V star (2.0 g cm$^{-3}$). 
This apparent 
discrepancy probably arises from the transit ingress and egress appearing less steep than 
expected for a 
main sequence star. To investigate the origin of this problem, we divided the LC  in 32 
groups, and individually fitted each group of transits (containing between 4 and 5 transits each). 
The fitted inclinations in groups of transits is systematically larger than the inclination obtained 
from the global phase-folded LC. The mean a/\rstar = 4.0 is significantly different and the resulting 
density 
(1.6 g cm$^{-3}$) agrees better with that of a G9V star. 

Consequently, we assume that the global transit gives slightly degraded information on the 
actual ingress and ingress of the transit. This may result in large errors on the inferred stellar  
parameters, including the density, if we  rely on the analysis with four free parameters.  We consider  two possible causes for this degradation.

The first one is transit timing variations (TTVs) -- temporal shifts to the centre  of each transit 
caused by the presence of  additional 
bodies in the system  \citep{2005MNRAS.359..567A,2005Sci...307.1288H}. In 
fact, when each group of transit is shifted accordingly to the best-fitted center, and the combined 
transit is built, we obtain a shape with a steeper ingress/egress, thus 
alleviating the discrepancy. However, the time scales of the putative TTVs and their amplitudes are 
not 
easily understood in terms of gravitational interactions with other bodies, due to the short distance 
between \corotsevenb\, and its host star. We thus favour a second explanation in terms of the stellar 
activity. Several works (\citealt{2008MNRAS.385..109P,2009IAUS..253...91A}) have shown that 
the occultations of active regions and/or spots can induce apparent shifts of the transit centers that 
might erroneously be attributed to additional bodies in planetary systems. Because the host star 
\corotseven\, is clearly an active star, and because of  the comparable sizes of the transiting object and 
the stellar spots, we suspect that this effect is important. Unfortunately,  to verify this 
hypothesis, we would need photometry of individual transits with the same order of precision as 
the combined transit (54~ppm per 1.5~min data point), which will be difficult to achieve in the next 
few years.

These limitations mean that  we are not able to obtain  altogether  precise stellar parameters and 
planetary
parameters from the LC alone, as is the case  for giant planets with a high S/N transit.
 The solution is too degenerate as illustrated by Fig. 
\ref{chi2-4paras}  where the contour map of chi2 is plotted in the [\rplanet\,, inclination ] frame: 
clearly the range of possible solutions within the 2 $\sigma$ contour is too broad to be useful; 
e.g. the resulting ranges for the planet radius and orbital inclination are \rplanet\, $\in$ [1.4 -- 2.3 
\RE] 
and $i  \in$ [60 -- 90$^{\rm o}$], taking the uncertainty on \rstar\,  into account and with a 5\% risk of 
error.

Instead, we rely on the spectroscopic analysis described in Sect. 7 and make the assumptions 
of  a circular orbit and of a limb-darkening law following \citet{2000A&A...363.1081C} quadratic 
approximation.  By forcing 
the stellar radius to be \rstar = 0.87$\pm$0.04 and considering the phase-folded light curve of Fig. 
\ref{LC_folded} (i.e. without any correction for putative TTVs), we looked for the best fit when 
leaving two free parameters: the radius of the planet and the inclination of the orbit. To 
assess the significance of the best fit and estimate uncertainties, we divided the light curve into five 
phase-folded subsets and computed the fitting parameters in each case. The mean and 
standard deviation were then computed for each of the derived parameters: planetary radius, 
transit duration, and inclination.

We obtain  \rplanet\, = 1.68 $\pm$ 0.09 \RE, T$_{14} = 1.125 \pm
$ 0.05 h, and $i$ = 80.1 $\pm$ 0.3$^{\rm o}$. The error bars are fully dominated by the 
uncertainties 
on the stellar parameters. The resulting transit curve (Fig. 
\ref{Corot_exo_0165_phasefolded_17feb09_v2})
can be compared to the observations. We must grant that the agreement is not fully satisfying,
especially in the ingress, but we also note equivalent  residual structures at phase values out of 
the transit
that may indicate the effect of stellar activity,  as previously stated.

From the previous analysis, the mean half-length of the transit projected on the stellar disk, in 
stellar radius unit, is h = ($\pi$ a $\tau$ / P)/ \rstar = 0.71 $\pm$ 0.06 , and the  impact parameter is 
0.70  
$\pm$ 0.06.
The final set of adopted planet parameters is summarised in Table \ref{table_planet}. 

\begin{table}
\begin{center}{
\caption{\label {table_planet} Planetary parameters.}
\begin{tabular}{lll}
\hline
\hline
    Parameter      & Value   &  Uncertainty   \\
\hline
Period (day)  &  0.853585 & $\pm$ 2.4\, 10$^{-5}$      \\   
a (AU)   &  0.0172   &  $\pm$  2.9\, 10$^{-4}$    \\  
a/\rstar  &   4.27  &  $\pm$ 0.20   \\  
T$_{14}$ (h) & 1.125   & $\pm$    0.05    \\ 
impact parameter z & 0.61 &  $\pm$ 0.06 \\
k = \rplanet\,  / \rstar &   0.0187 & $\pm$  3\, 10$^{-4}$   \\  
\rplanet\,  / \RE & 1.68  & $\pm$  0.09 \\
\mplanet / \ME & $<$ 21  & \\
$i$ (deg) &   80.1  & $\pm$ 0.3   \\
\hline
\end{tabular}}
\end{center}
\end{table}

\section{Discussion}
\label{sec:discussion}.
\subsection{Tidal and centrifugal force effects}

The star and the planet are exchanging strong tidal forces. Tidal forces influence the motion and 
the evolution the Corot-7 system. One consequence is the planetary spin-orbit coupling. 
According to 
\cite{1999ssd..book.....M}, the star raises tides on the planet that lead to the synchronization of the 
planetary rotation with its revolution, in a characteristic time $\tau_{synch}$ 

\begin{math}
\tau_{synch} = \frac{ |(n - \Omega_{p}| }  {  \frac{3}{2}  \frac{\mmstar }{\mmplanet }  (\frac{\mrplanet }
{a})^3 (\frac{G \mmstar }{a^3})  }  I \frac{Q_{p}}{k_{2p}}
\end{math},
\\ 
where $n$ is the mean motion of the revolution rate of the planet,  $\Omega_{p}$ the primordial 
rotation rate of the planet, $I$ the normalized moment of inertia of the planet, $Q_{p}$ the 
planetary dissipation constant, and $k_{2p} $ the Love number of second order. Several of the 
stellar and planetary parameters have been determined in this work. Some planetary 
characteristics are unknown or poorly known, but can be estimated within reasonable ranges. The 
normalized moment of inertia $I$ describes how the mass of a body is distributed in its interior. If 
the body is differentiated (a safe assumption for a body larger than 1000 km in size), the heavier 
materials are concentrated in the core and $I <$ 0.4. The planets and even the large moons of the 
Solar System have 0.2 $< I < $ 0.35.    The solar terrestrial planets show values of the Love 
number $Q_{p} / k_{2p}$ between 30 and 1000 \citep{1995geph.conf....1Y}. The primordial 
planetary rotation rate $\Omega_{p}$ is not known, of course, but using values for a fast rotator (10 
hours) and a slow rotator (10 days) and using the estimated parameters in the ranges above 
yields a time constant $\tau_{synch}$ \textit{in the range one year to decades}. 

As a consequence, regardless of the poorly known planetary parameters, the synchronization of 
the planetary 
rotation with its revolution is a fast and efficient process that has  already been completed, given 
that  the age of system 
age (1 --  2 Gyr) is much longer than $\tau_{synch}$. Although the planet has a telluric nature and 
the planetary to stellar mass ratio is 
small, the decisive factor for the efficiency of tidal dissipation is the short distance to the star.

The stability of the planetary orbit under the influence of tidal forces depends crucially on the ratio 
of the stellar dissipation rate and the stellar Love number $Q^*/k_{2}^*$ 
\citep{2007P&SS...55..643C}. 
The time scale for the decay of the planetary orbit from the currently observed distance toward the 
Roche zone of the star is \citep{2002ApJ...568L.117P,2004A&A...427.1075P}

\begin{math}
 \tau = \frac{ \frac{2}{13} [a^{12/3} - a_{Roche}^{12/3}]  Q_{\star} }
 { 3 (\mmplanet / \mmstar) \mrstar^{5} (G \mmstar)^{1/2} k_{2\star} } 
\end{math}, \\
with $a_{Roche} = 2.46 \mrstar$ as the Roche radius of the star \citep{1987QB410.C47......}. The 
dissipation constant and the Love number of a star are poorly known. Values for $Q_{\star}/k_{2 
\star}$ 
vary in wide ranges in the literature. Values from 10$^5$ to 10$^{5.5}$ 
\citep{1996Natur.380..606L,2008ApJ...678.1396J} 
would yield unrealistic small time scales of 70 Myr for the decay because it would be highly 
improbable to observe this planet today. For $Q_{\star}/k_{2 \star}$  = 10$^6$ to 10$^{6.5}$, the 
orbit would decay within 2 Gyr. The orbit may be considered stable with respect to tidal forces for 
$Q_{\star}/k_{2 \star} > 10^7$. The latter limit has also been derived by 
\cite{2007P&SS...55..643C} for the case of OGLE-TR-56b and seems to fit observations  better. These values were computed with the upper 
planetary mass limit of 21 Earth masses.

The shape of the planet is a triaxial Roche 
ellipsoid \citep{1987QB410.C47......}, distorted by the tidal and rotational potentials. 
The longest of the ellipsoid axes is directed towards the star 
while the shortest is directed along the rotation axis of the planet. In the case of an homogeneous 
distribution of mass, the equator prolateness (tidal bulge) of the Roche ellipsoid is given by 
$(15/4)(\mmstar/\mmplanet)(\mrplanet/a)^3$, and the polar flattening, referred to the mean 
equatorial radius, by  $(25/8)(\mmstar/\mmplanet)(\mrplanet/a)^3$. Using the values given in Table 
\ref{table_planet} and Ê\mplanet $<$ 21 \ME, we obtain an equator prolateness $<$  0.016 and a 
polar flattening $<$ 0.013. If the non-uniform distribution of masses in the interior of the planet is 
taken 
into account, the results are smaller. If we use the reduction factor that corresponds to the Earth ($
\approx$ 0.78), the expected values are 0.0125 ($\approx 1/80$) for the polar flattening and  
0.010 ($\approx 1/100$) for the tidal bulge. This means that the largest equator radius of the 
planet (high 
tide) is $<$ 140 km larger than the shortest equator radius (low tide) and that the polar radius is 
$ <$120 km smaller than the mean equatorial radius. 

 In any case, the corresponding stretching of the planet is small enough to be neglected in the 
estimate of its volume because the corresponding uncertainty is less than on the radius 
determination from the transit and spectroscopy ($\pm$ 570 km).

\subsection{ Composition of the planet}

  \begin{figure}
   \centering
\includegraphics[width=8.5cm]{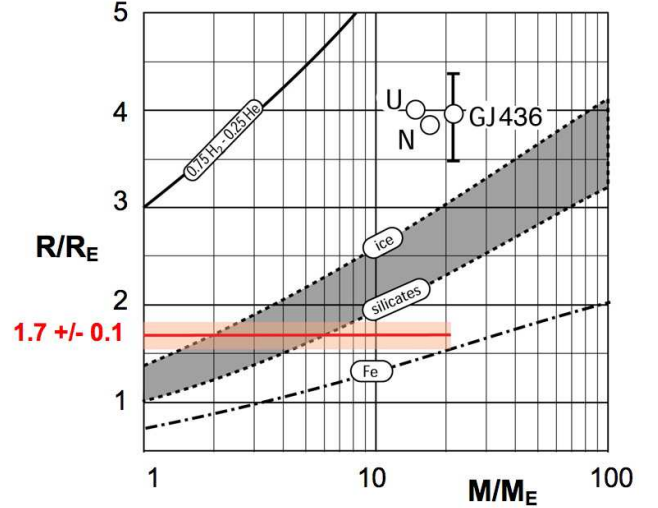}
      \caption{
 Planetary radius as a function of mass for different compositions of planets 
\citep{2009arXiv0902.1640G}. The curves [Fe], [ silicates], [ices], [H$_2$-He] correspond to 
planets made of pure Fe, silicates, and metallic core (analogous to the Earth), pure water ice and 
pure H$_2$-He gas, respectively. The shaded area corresponds to planets with both silicates and 
water. The 
region between this area and the curves [H$_2$-He] corresponds to planets with a water-silicate 
core and a thick H$_2$-He envelope, e.g. Uranus, Neptune or GJ 436. 
The red band corresponds to the present determination of the radius, \rplanet\,  = 1.68 $\pm$ 0.09 
\RE, and upper limit for the mass, \mplanet $<$ 21 \ME. A purely iron planet can be excluded.
 \label{radius-vs-mass}
  }
   \end{figure}

The estimated radius (1.68 $\pm$ 0.09 \RE ) and mass limit ($<$ 21 \ME) of the planet can be 
located in 
the (R, M) plane and compared with the R(M) relations derived from the modellings of the internal 
structures of different planets \citep{2007Icar..191..337S,2008ApJ...688..628E,
2007ApJ...665.1413V,2009arXiv0902.1640G}. As shown in Fig.\ref{radius-vs-mass}, the present 
constraints are not 
strong and only exclude \corotseven\,  being a purely metallic planet. 

According to \cite{Lammer09}, hydrogen, if present when the planet was formed, would be 
driven away by thermal and non-thermal processes (Jeans escape from the exosphere plus 
sputtering 
and ion exchange with the stellar wind) in a time much shorter than the system age ( $>$ 1 Gyr). 
In the absence of hydrogen, the main components of the planet can be water, silicates, and metals. 
If the planetary mass can be determined more precisely, a better determination of the 
composition is possible. The ambiguity between a rocky planet and one containing a 
significant amount of light elements could be overcome. 

If the preliminary result, 5 Mearth $<$ \mplanet $<$ 11 \ME that we have obtained at the time of 
submitting the present paper and announced at the CoRoT Symposium 2-5 February 2009 in Paris, is confirmed, it 
would point to, or at least be compatible with, a rocky planet  (Fig. \ref{radius-vs-mass}). 

\subsection{Temperature at the planetary surface}
 
 The planet is very close to a G9V star  (a = 0.0172 AU $\pm$ 0.0003 = 4.27 $\pm$ 0.20  \rstar\,),
and its spin and orbital rotations are most likely phase-locked. The stellar disk is seen from the 
peristellar point is enormous, 28$^{\rm o}$ in diameter. A high temperature is then expected at 
the surface on the day side of the planet. However, an estimate of the temperature distribution 
depends upon different hypotheses, depending on whether or not an efficient mechanism 
for transferring the energy from the day side to the night side is present.\\
-- If such a mechanism exists and the temperature is almost uniform as on Venus, it would be\\
(\tplanet)$_1$ = (1 - A)$^{1/4} {\mathcal G} (\mrstar/ 2a)^{1/2}$ \tstar , \\
where $A$ is the planetary albedo and ${\mathcal G}$  stands for the greenhouse effect. 
Assuming $A$ = 0 and ${\mathcal G}$ = 1  for a rocky planet without an atmosphere, the temperature reads 
(\tplanet)$_1$ = 1810 $\pm$ 90  K,
the uncertainty on T reflecting those on \tstar and a/\rstar. \\
-- If there is no such mechanism, the temperature is the result of the local balance 
between impinging and emitted powers. In the (crude) approximation where the incident light 
beam 
from the star is parallel, the temperature is\\
(\tplanet)$_2 \approx$ (1 - A)$^{1/4}$ ${\mathcal G}$ (\rstar/ a)$^{1/2}$ \tstar $(\cos \Phi)^{1/4}$ , 
for  $\Phi  \in 
[0^{\rm o},  90^{\rm o}]$,\\
where $\Phi$ is the angle between the normal at the surface and the planet-star axis ($\Phi$  = 0 
at the substellar point, and $\Phi$ = 90$^{\rm o}$ at the terminator).
At the substellar point, for $A$ = 0 and ${\mathcal G}$ = 1, the temperature reads (\tplanet ($\Phi = 
0$))$_2$ = 2560 $\pm$ 125 K. 
 
In the latter  hypothesis and in the absence of an atmosphere that produces a Greenhouse effect, 
\textit{the temperature of the night side, i.e.  $\Phi  \in 
[90^{\rm o},  180^{\rm o}]$, can be surprisingly low} because it mainly faces the cold outer 
space. This situation is similar to that of the north and south poles of the Moon  ($\approx$  40 K), 
and dark 
face of Mercury ($\approx$ 90 K) \citep{1999Icar..141..179V}. If a geothermal flux of 300 mW m
$^{-2}$ is the main heating process, the temperature 
would be $\approx$ 50 K. 

\section{Conclusions}

The CoRoT satellite has discovered transits around the star \corotseven\, that are compatible 
with the presence of a small planet. Using ground-based follow-up observations and the 
satellite colour light-curves, we discarded almost all conceivable cases of false positives. 
 In so far as we have been exhaustive in listing the cases of these possible 
false positives, we conclude that we have discovered the smallest exoplanet known to date, with a 
radius \rplanet\, = 1.68 $\pm$ 0.09 \RE. Taking into account the possibility of a chance alignment 
of a BEB at less than 400 mas from the target that is not excluded by our 
follow-up, the actual presence of this planet can be considered as established with a risk of  a
false positive conservatively estimated to 8 10$^{-4}$. 

The amplitude of transits is $\Delta F / F \approx 3.35 10^{-4}$ $\pm$ 0.12 10$^{-4}$ (trapezoidal 
approximation), as detected by the 
satellite. The star is characterized with high-resolution spectroscopy and is considered 
as an active star with spectral type  G9V. At the date of this paper's submission (Feb. 2009), the 
information on the planetary mass resulting from RV measurements is only an upper limit, 
\mplanet $<$ 21 \ME. The planetary orbital period, 0.8536 days, is the shortest one ever detected 
(http://exoplanet.eu). The corresponding proximity of the planet to its star (a = 0.0172 AU = 
4.3 \rstar) implies a high temperature at its surface. At the substellar point, assuming a 
zero albedo and no Greenhouse effect, it is $\approx$ 1800 K to 2600 K, depending on whether there 
is an efficient redistribution of the energy on the planetary surface.

Taking the preceding reserves into account , it should be noted that it is possible to 
deduce the presence of a small orbiting planet with only a small risk of false detection 
($<$ 8 10$^{-4}$) without a formal RV detection. Even the information that 
there is neither a Jupiter mass planet nor a stellar companion around the main target star
is provided by the duration of the transit, e.g. a grazing Jupiter would give a shorter transit 
than what is observed. To our present knowledge, a 1.68 Earth radius object can only
be  a telluric planet or a white dwarf. Because the latter case is easily discarded by the RV 
measurements since it would lead to a very large signal, we conclude  that there is 
a telluric planet. This situation will probably repeat in the future, e.g. when the results 
from the Kepler mission come, as the search for 
habitable terrestrial planets becomes a central scientific issue and the confirmation by RV 
very difficult. For RV measurements, \corotsevenb\, is a favourable case because the expected 
signal is stronger than for similar planets in the HZ, and its short period allows the study of many 
orbits 
during a given duration of the observations, e.g. over 100 orbits during 4 months. It can be 
noted that, if this planet had  one Earth mass and was in the HZ of its star (orbital period of $
\approx$ 
220 days), the amplitude of the RV reflex motion of the star would be $\approx$ 130 times smaller 
than 
what can be presently excluded (k $<$ 15 m/s); the confirmation by RV would then have probably been impossible in the present state of the art. 

If the presently ongoing efforts in RV measurements on CoRoT-7 are successful, they would 
be very valuable because they would allow: (i) an independent detection of the planet, (ii) a 
determination of its mass. The latter information would permit precise inferences on its 
composition, possibly including  between a rocky and a water-rich planet.

\begin{acknowledgements}
The authors are grateful to all the people that have worked on and operated 
the \corot\, satellite, including Adam C., Ballans H., Barbet D. , Bernard M., Collin C., Docclo A., 
Dupuis O., Essasbou H., Gillard F., GuŽriau A-C., Joguet M., Levieuge B., Oulali A., Parisot J., Pau 
S., Piacentino A., Polizzi D., Reess J-M., Rivet J-P., Semery A., Strul D., Talureau B. They are 
grateful to Despois, D., Selsis F.,  Zuckerman B. for stimulating discussions.
 HJD and JMA acknowledge support by grants ESP2004-03855-C03-03 and
ESP2007-65480-C02-02 of the Spanish Education and Science Ministry.
RA acknowledges support by grant  CNES-COROT-070879. 
The German \corot\, Team (TLS and Univ. Cologne) acknowledges DLR
grants 50OW0204, 50OW0603, 50QP07011. The building of the input \corot/Exoplanet catalogue
was made possible thanks to observations collected for years at the Isaac Newton Telescope 
(INT), operated on the island of La Palma by  the Isaac Newton group in the Spanish Observatorio 
del Roque de Los Muchachos of the Instituto de Astrofisica de Canarias. We thank  R. Rebolo and the FastCam teams at IAC and UPCT for 
permission to use their camera during technical testing time.
The authors are also grateful to an anonymous referee who helped in improving the 
manuscript significantly. 
\end{acknowledgements}

\bibliographystyle{aa}
\bibliography{E2_165}

\end{document}